\documentclass{jpsj2}
\usepackage{graphicx}
\usepackage{graphics}
\usepackage{amssymb}

\newcommand{\bb}{\mbox{\boldmath $b$}}

\newcommand{\sgn}{ ~{\rm sgn} }

\newcommand{\be}{\begin{equation}}
\newcommand{\ee}{\end{equation}}
\newcommand{\bd}{\begin{displaymath}}
\newcommand{\ed}{\end{displaymath}}
\newcommand{\bra}{\langle}
\newcommand{\ket}{\rangle}

\newcommand{\rmi}{\mathrm{i}}
\newcommand{\rmd}{\mathrm{d}}
\newcommand{\rme}{\mathrm{e}}

\renewcommand{\bb}{\mbox{\boldmath $b$}}
\newcommand{\bs}{\mbox{\boldmath $s$}}
\newcommand{\Bra}{\left\langle}
\newcommand{\Ket}{\right\rangle}

\def\runauthor{Online-Journal Subcommittee}

\title{%
Counting solutions for the CDMA multiuser MAP demodulator }

\author{%
J.P.L. Hatchett \dag \thanks{E-mail address :
Jon.Hatchett@hymans.co.uk} and Jun-ichi Inoue \ddag }

\inst{%
\dag Hymans Robertson LLP, One London Wall, London EC2Y 5EA, UK \\
\ddag Complex Systems Engineering,
Graduate School of Information Science and
Technology, Hokkaido University,
N14-W9, Kita-ku, Sapporo, 060-0814, Japan
}

\recdate{\today}

\abst{%
We evaluate the average number of locally minimal solutions for
maximum-{\it a-posteriori} (MAP) demodulation in code-division
multiple-access (CDMA) systems. For this purpose, we use a
sophisticated method to investigate the ground state properties for
the Sherrington-Kirkpatrick-type (i.e. fully connected) spin glasses
established by Tanaka and Edwards in 1980. We derive the number of
locally minimal solutions as a function of several parameters which
specify the CDMA multiuser MAP demodulator. We also calculate the
distribution function of the normalized-energies for the locally minimum
states. We find that for a small number of chip intervals (or
equivalently a large number of users) and large noise level at the
base station, the number of local minimum solutions becomes larger
than that of the SK model. This provides us with useful information
about the computational complexity of the MAP demodulator. }

\kword{%
CDMA, Statistical Mechanics, Spin Glasses,
Metastable States, Bayesian Statistics,
Tanaka-Edwards Theory, Replica method
}

\begin{document}
\maketitle
\section{Introduction}

Recently, statistical-mechanical analysis has revealed many
important aspects of probabilistic information
processing\cite{Nishimori}. Among them, many studies addressing the
code-division multiple-access (CDMA) communication problem succeeded
not only in investigating the statistical properties of demodulators
\cite{Tanaka1,Tanaka2} but also constructing iterative algorithms
based on so-called belief-propagation\cite{TanakaOkada, Kabashima}
and examining the dynamics of decoding algorithms\cite{MimuraOkada,
HatchettOkada}. Within the framework of Bayesian inference,
marginal-posterior-mode (MPM) demodulation provides the best
possible performance in the sense that its bit-error rate is
minimized under a specific condition, namely, the so-called
Nishimori condition\cite{Tanaka1,Tanaka2}. However, it is also
possible for us to choose another strategy, that is, the
maximum-{\it a-posteriori} (MAP) demodulator, in order to estimate
the original information bit for each user from the received signals
at the base station. The MAP demodulator attempts to achieve the
maximum in the posterior distribution of the sent bits. In practice
this means that we choose as our estimate of the original
information bits the ground state of the Hamiltonian, which is
defined by minus the logarithm of the posterior distribution.
Therefore, it is quite important for us to be able to evaluate how
many local minimum (metastable) solutions exist around the ground
state, as when we attempt to minimize the Hamiltonian to obtain the
ground state many local search algorithms will get trapped in these
metastable states. 
However, 
there are few studies to investigate such a 
computational complexity 
aspect of the CDMA multiuser demodulator. 
For this kind of problem, Tanaka and
Edwards\cite{TE,TE2} established a general theory to count the number of
locally minimum energy states for the model class of the
Sherrington-Kirkpatrick spin glasses\cite{SK}. They showed that the
average number of the local minimum states scales (to leading order
in $N$) as $ \sim {\rm e}^{0.19923N}= 2^{0.28743N}$ for the SK-type
long-range mean-field model. As there exists a close relationship
between the CDMA multiuser demodulator and the Hopfield model with
extensive number of patterns, which is itself strongly disordered in
a similar way to the SK model, their method appeared to be useful
for us to investigate the ground state properties of the CDMA
multiuser MAP demodulator. With the assistance of the Tanaka-Edwards
theory, in this paper, we evaluate the number of the locally minimum 
normalized-energy solutions for CDMA multiuser MAP demodulation problems. 
We also calculate the distribution of the normalized-energies 
of the local minimum 
states and discuss how often 
we obtain the deep local minimum energy 
level of the solution for a given parameter set, namely, 
the number of users and the noise level at the base station.

This paper is organized as follows. In the next section, following
the scheme of Tanaka\cite{Tanaka1,Tanaka2}, we introduce a model
system for the CDMA multiuser demodulator. The MAP demodulator is
formulated in the context of Bayesian statistics. Then, the energy
function to be minimized is introduced naturally. In the same
section, we define the local minimum solution of the MAP demodulator
as a fixed point of the zero-temperature dynamics of the CDMA
multiuser demodulation problem. 
In section 3, we evaluate the number 
of local minimum solutions for the MAP demodulator by using 
Tanaka-Edwards theory\cite{TE,TE2}. In the following section, we 
calculate the distribution of the local minimum normalized-energies 
and make it clear how often the deep locally minimum state appears for a given 
number of users and noise level at the base station. It is 
well-known that the Tanaka-Edwards theory is based on the annealed 
calculation for the average of the macroscopic quantities of the 
system. Therefore, it cannot take into account the effect of the 
quenched disorder correctly. In order to evaluate the number of 
local minimum states in the proper way in section 5 we attempt to 
recalculate the number of the locally minimum solutions for the MAP 
demodulator modeling it as a quenched system with assistance of the 
replica method. Then, we compare the result of the annealed 
calculation with that calculated by the replica method. The last 
section is devoted to a summary of our results.

\section{Bayesian inference for a model CDMA system}
We consider the direct-sequence binary phase-shift-keying (DS/BPSK)
CDMA code where we have $K$ users, $s_1^{0},\ldots,s_K^{0}$ which
use a spreading code $b_k^1,\ldots,b_k^N$, where we define $N/K
\equiv \alpha$. The received signal is given by
\begin{eqnarray}
y^\mu = \frac{1}{\sqrt{K}} \sum_{k=1}^K b_k^\mu s_k^{0} + \nu^\mu,
\label{eq:true_system}
\end{eqnarray}
where  $\nu^\mu \sim N(0,1/\beta_s)$ is Gaussian noise and the label
$\mu$ takes the values $\mu=1,\cdots,N$. 
$N$ is the number of 
components of spreading codes for each user. 
The CDMA demodulation
problem is to estimate the original information for each user
$s_1^{0},\ldots,s_K^{0}$ given that the output $y^{t}$ and the
spreading code for each user $b_k^1,\ldots,b_k^N$ is known. We model
the spreading code sequences $\{b_k^\mu\}$ as sequences of
independent identically distributed binary random variables with
Prob[$b_i^{\mu} = \pm 1$] = 1/2.

For this problem, we introduce a model of the system
(\ref{eq:true_system}):
\begin{eqnarray}
y^\mu = \frac{1}{\sqrt{K}} \sum_{k=1}^K b_k^\mu s_{k}
+ \bar{\nu}^\mu,
\end{eqnarray}
where $s_1,\ldots,s_K$ are estimates of the corresponding original
information bit $s_1^{0},\ldots,s_K^{0}$ for each user and
$\bar{\nu}^t$ is a model of the noise at the base station which
follows a Gaussian distribution with variance $\beta^{-1}$.

Within the context of Bayesian statistics, we consider the posterior
distribution which is given by
\begin{multline}
P(\{s_{k}\}| \{y^{\mu}\}, \{b_{k}^{\mu}\}) =
\frac{\beta}{\sqrt{2\pi}} {\exp} \left[ -\frac{\beta}{2}
\sum_{\mu=1}^{N} \left( y^{\mu}- \frac{1}{\sqrt{K}} \sum_{k=1}^{K}
b_{k}^{\mu} s_{k} \right)^{2}
\right] \\
=
\frac{\beta}{\sqrt{2\pi}}
{\exp}
\left[
-\frac{\beta}{2}
\sum_{\mu=1}^{N}
(y^{\mu})^{2}-
\beta H(\bs)
\right], 
\end{multline}
where we 
choose the uniform distribution 
$P(\{s_{k} \})=2^{-K}$ as a prior. 
Using these definitions we are working with a system which has an
effective Hamiltonian given by
\begin{eqnarray}
H(\bs) = \frac12 \sum_{i,j} s_i J_{ij} s_j - \sum_i f_i s_i
\label{eq:hamilton}\\
J_{ij} = \frac1K \sum_{\mu = 1}^N b_i^\mu b_j^\mu\qquad f_k =
\frac{1}{\sqrt{K}} \sum_{\mu =1}^N y^\mu b_k^\mu,
\end{eqnarray}
which is constructed so that maximizing the posterior distribution
corresponds to minimizing this Hamiltonian (\ref{eq:hamilton}). The
Hamiltonian gives rise to local fields acting on each spin $s_k$
\begin{eqnarray}
h_k(\bs) = f_k - \sum_{j \neq k} J_{kj} s_j
\label{eq:local_field}
\end{eqnarray}
so that the zero temperature dynamics, which attempts to find the
maximum-{\it a-posteriori} (MAP) demodulator solution, is given via
\begin{eqnarray}
s_k = \sgn(h_k(\bs)). \label{eq:zero_T_dynamics}
\end{eqnarray}
Additionally, this means that a given state $\bs$ is a stable fixed
point of the dynamics if, and only if, $h_k(\bs) = \lambda_k s_k,\
\lambda_k \geq 0\ \forall k$.
We would like to comment that the above estimate becomes the
so-called conventional demodulator (CD) if we neglect the second
term of the right hand side of equation (\ref{eq:local_field}) (i.e.
we ignore the interactions between different bits in the received
message).

The dynamics (\ref{eq:zero_T_dynamics}) actually minimizes the
Hamiltonian (\ref{eq:hamilton}) (i.e. finds its global minimum) if
there is no local minimum in the energy landscape. However, due to
the quenched disorder $\{b_{k}^{\mu}\}$ in (\ref{eq:hamilton}),
which manifests itself in the interactions $J_{ij}$ and fields
$f_{k}$, there exists many local minima and the dynamics
(\ref{eq:zero_T_dynamics}) may well become trapped in one of the
locally minimum states. Therefore, it is quite important for us to
evaluate how many locally minimum states exist around the globally
minimum state. Obviously, the number of solutions depends on the
number of users $K$ and the noise level at the base station
$\beta_{s}$. Our main goal in this study is to make this point clear
in a quantitative manner.

\section{The average number of local minimum states : annealed calculation}
In this section, following the method developed by Tanaka and
Edwards\cite{TE,TE2}(especially, by their formulation 
for the Ising case \cite{TE}), we calculate the number of solutions to the
dynamical equations (\ref{eq:zero_T_dynamics}), which are attempting
to construct the MAP demodulator 
for a given solution $\mbox{\boldmath $s$}$ 
of the dynamics (\ref{eq:zero_T_dynamics}). 
We first define the local energy
$\epsilon_{i}$ by
\begin{eqnarray}
\epsilon_{i} = -s_{i} f_{i} +  s_{i} \sum_{j \neq i} J_{ij} s_{j}.
\end{eqnarray}
Then, 
according to Tanaka and Edwards \cite{TE}, 
we assume that 
each bit asynchronously updates and 
the energy difference due to the bit flip $s_{i} \to -s_{i}$
is given by
\begin{eqnarray}
\Delta \epsilon_{i} = \epsilon_{i}^{'} - \epsilon_{i} = 2s_{i}
\left( f_{i}-\sum_{j \neq i} J_{ij} s_{j} \right).
\end{eqnarray}
If we define the parameter $\lambda_{i}$ in terms of the local field
$h_{i}(\bs)$ via
\begin{eqnarray}
h_{i}(\bs) = f_{i} - \sum_{j \neq i} J_{ij}s_{j} = \lambda_{i}s_{i},
\end{eqnarray}
then 
for a given realization of disorder $\{J,f\}$, 
the condition for the solution $\bs$ to be one of the locally
minimum states is given as $\Delta \epsilon_{i} >0\ \forall i$,
namely,
\begin{eqnarray}
\Delta \epsilon_{i} = 2s_{i} \lambda_{i} s_{i} =2\lambda_{i} >0
\label{eq:cond_stable}.
\end{eqnarray}
Of course, 
we might modify the condition 
(\ref{eq:cond_stable}) to investigate the stability against 
a cluster spin flip, however, 
the analysis for such cases is beyond the scope of 
our present abilities. 

Therefore, we can calculate the average number of locally minimum
states, $\bra g_0 \ket$, through
\begin{eqnarray}
\bra g_0 \ket =
\Bra \sum_{\mbox{\boldmath $s$}}
\prod_i
\Theta (\lambda_{i}) \Ket_{\{J,f\}} =
\Bra \sum_{\mbox{\boldmath $s$}}
\prod_i \left[ \int_0^{\infty} \rmd
\lambda_i \delta \left(f_i - \sum_{j \neq i} J_{ij} s_j - \lambda_i
s_i \right) \right] \Ket_{\{J,f\}}
\end{eqnarray}
where $\Theta(\cdots)$ denotes the step function. 
The $f_k$ are defined in terms of 
the output of the Gaussian channel $y^\mu$, the measure of which is
given by the probability measure
\begin{eqnarray}
Z(\mathbf{y}) = 2^{-K} \left(\frac{\beta_s}{2\pi}
\right)^\frac{N}{2} \sum_{\bs^0} \exp\left[-\frac{\beta_s}{2}
\sum_{\mu = 1}^N \left(y^\mu - \frac{1}{\sqrt{K}} \sum_{k = 1}^K
b_k^{\mu} s_k^0 \right)^2 \right].
\label{eq:partition}
\end{eqnarray}
Thus we may write the average number of fixed points of the
dynamics (\ref{eq:zero_T_dynamics}) as
\begin{multline}
\bra g_0 \ket = 2^{-K}\left(\frac{\beta_s}{2\pi}
\right)^\frac{N}{2} \sum_{\bs^0} \int \prod_{\mu = 1}^N \rmd y^\mu
\exp\left[-\frac{\beta_s}{2} \sum_{\mu = 1}^N \left(y^\mu -
\frac{1}{\sqrt{K}} \sum_{k = 1}^K b_k^\mu s_k^0 \right)^2 \right] \\
2^{-NK} \sum_{\bs, \bb^1,\ldots,\bb^N} \prod_i \int_0^\infty \rmd
\lambda_i \int_{-\rmi\infty}^{\rmi\infty} \frac{\rmd
\hat{\lambda}_i}{2\pi \rmi} \exp\left[\hat{\lambda}_i
\left(\frac{1}{\sqrt{K}} \sum_\mu y^\mu b_i^\mu - \frac{1}{K}
\sum_{j \neq i} \sum_\mu b_i^\mu b_j^\mu s_j - \lambda_i s_i \right)
\right] \label{eq:anneal_g0}
\end{multline}
under the assumption that the system is well-approximated as an
annealed system. By using the saddle point method in the limit of $K
\to \infty$, the average $\bra g_{0} \ket$ under the annealed
approximation is given by
\begin{multline}
\bra g_0 \ket=\int \frac{\rmd t \rmd \hat{t}}{2\pi/K} \frac{\rmd u
\rmd \hat{u}}{2\pi/K} \frac{\rmd w \rmd \hat{w}}{2\pi/K} \frac{\rmd
q \rmd \hat{q}}{2\pi/K} \rme^{ K \Phi(
t,u,w,q,\hat{t},\hat{u},\hat{w},\hat{q})} \\
\Phi = \hat{t}t + \hat{u} u + \hat{w} w + \hat{q}q
-\frac{\alpha}{2} \log \left\{ u[1 + 2\beta_s(1-q)] + \beta_s(1 + t
-w)^2\right\} \\
+\log \left\{\cosh(\hat{q}) + \frac12 \left[\rme^{-\hat{q}}
\mbox{Erf} \left(\frac{\alpha -\hat{t} - \hat{w}}{2\sqrt{\hat{u}}}
\right) + \rme^{\hat{q}} \mbox{Erf} \left(\frac{\alpha -\hat{t} +
\hat{w}}{2\sqrt{\hat{u}}} \right) \right]\right\} + \frac{\alpha}{2}
\log \beta_s
\label{eq:anneal_g0_2}.
\end{multline}
The details of the derivation is shown in Appendix A.

We now have to vary this saddle point surface to find its extremum.
The definition we have used for the error function is Erf$(z) =
(2/\sqrt{\pi}) \int_0^z \rmd t \,\rme^{-t^2}$ so that
\begin{eqnarray}
\frac{\rmd}{\rmd z} \mbox{Erf}(z) = \frac{2}{\sqrt{\pi}} \, \rme^{-z^2}.
\end{eqnarray}
Varying with respect to the true parameters gives
\begin{subequations}
\begin{align}
\hat{t} &= \frac{\alpha \beta_s (1 +t -w)}{u[1 + 2\beta_s(1-q)] +
\beta_s(1 + t -w)^2}\\
\hat{w} &= -\hat{t}\\
\hat{u} &= \frac12\frac{\alpha[1 + 2\beta_s(1-q)] }{u[1 +
2\beta_s(1-q)] +
\beta_s(1 + t -w)^2}\\
\hat{q} &= \frac{-\alpha u \beta_s}{u[1 + 2\beta_s(1-q)] +
\beta_s(1 + t -w)^2}
\end{align}
\end{subequations}
while varying the conjugate parameters leads to
\begin{subequations}
\begin{align}
q &= \frac{ \frac12
  \left[\rme^{-\hat{q}} \mbox{Erf}
\left(\frac{\alpha -\hat{t} - \hat{w}}{2\sqrt{\hat{u}}} \right) -
\rme^{\hat{q}} \mbox{Erf} \left(\frac{\alpha -\hat{t} +
\hat{w}}{2\sqrt{\hat{u}}}
\right) \right]-\sinh(\hat{q})}{\cosh(\hat{q}) + \frac12
  \left[\rme^{-\hat{q}} \mbox{Erf}
\left(\frac{\alpha -\hat{t} - \hat{w}}{2\sqrt{\hat{u}}} \right) +
\rme^{\hat{q}} \mbox{Erf} \left(\frac{\alpha -\hat{t} +
\hat{w}}{2\sqrt{\hat{u}}}
\right) \right]} \\
t &= \frac{\frac1{2\sqrt{\pi \hat{u}}}\left[\rme^{-q-[\frac{\alpha -\hat{t}
-
\hat{w}}{2\sqrt{\hat{u}}}]^2} +  \rme^{q-[\frac{\alpha -\hat{t} +
\hat{w}}{2\sqrt{\hat{u}}}]^2}\right]}{\cosh(\hat{q}) + \frac12
\left[\rme^{-\hat{q}} \mbox{Erf} \left(\frac{\alpha -\hat{t} -
\hat{w}}{2\sqrt{\hat{u}}} \right) + \rme^{\hat{q}} \mbox{Erf}
\left(\frac{\alpha
-\hat{t} + \hat{w}}{2\sqrt{\hat{u}}} \right) \right]}\\
w &= \frac{\frac1{2\sqrt{\pi \hat{u}}}\left[\rme^{-q-[\frac{\alpha -\hat{t}
-
\hat{w}}{2\sqrt{\hat{u}}}]^2} -  \rme^{q-[\frac{\alpha -\hat{t} +
\hat{w}}{2\sqrt{\hat{u}}}]^2}\right]}{\cosh(\hat{q}) + \frac12
\left[\rme^{-\hat{q}} \mbox{Erf} \left(\frac{\alpha -\hat{t} -
\hat{w}}{2\sqrt{\hat{u}}} \right) + \rme^{\hat{q}} \mbox{Erf}
\left(\frac{\alpha
-\hat{t} + \hat{w}}{2\sqrt{\hat{u}}} \right) \right]}\\
u &= \frac{\frac1{4\sqrt{\pi}\hat{u}^{\frac32}}\left[(\alpha - \hat{t}
- \hat{w})\rme^{-q-[\frac{\alpha -\hat{t} -
\hat{w}}{2\sqrt{\hat{u}}}]^2} + (\alpha - \hat{t} +
\hat{w})\rme^{q-[\frac{\alpha -\hat{t} +
\hat{w}}{2\sqrt{\hat{u}}}]^2}\right]}{\cosh(\hat{q}) + \frac12
\left[\rme^{-\hat{q}} \mbox{Erf} \left(\frac{\alpha -\hat{t} -
\hat{w}}{2\sqrt{\hat{u}}} \right) + \rme^{\hat{q}}  \mbox{Erf}
\left(\frac{\alpha
-\hat{t} + \hat{w}}{2\sqrt{\hat{u}}} \right) \right]}
\end{align}
\end{subequations}
We see that we can reduce the complexity slightly by introducing $s
\equiv t - w$ and $\hat{s} \equiv \hat{t} = -\hat{w}$ and then
eliminate $\{t,w, \hat{t}, \hat{w}\}$ in favour of $\{s,\hat{s}\}$.
This gives us the (slightly reduced) saddle point surface as
\begin{multline}
\Phi = \hat{s}s + \hat{u} u + \hat{q}q -\frac{\alpha}{2} \log
\left\{ u[1 + 2\beta_s(1-q)] + \beta_s(1 + s)^2\right\}\nonumber\\
+\log \left\{\cosh(\hat{q}) + \frac12 \left[\rme^{-\hat{q}}
\mbox{Erf} \left(\frac{\alpha}{2\sqrt{\hat{u}}} \right) +
\rme^{\hat{q}} \mbox{Erf} \left(\frac{\alpha
-2\hat{s}}{2\sqrt{\hat{u}}} \right) \right]\right\} +
\frac{\alpha}{2} \log \beta_s
\end{multline}
and the saddle point equations as:
\begin{subequations}
\begin{align}
\hat{s} &= \frac{\alpha \beta_s (1 +s )}{u[1 + 2\beta_s(1-q)] +
\beta_s(1 + s)^2}
\label{eq:hs} \\
\hat{u} &= \frac12\frac{\alpha[1 + 2\beta_s(1-q)] }{u[1 +
2\beta_s(1-q)] +
\beta_s(1 + s)^2}
\label{eq:hu} \\
\hat{q} &= \frac{-\alpha u \beta_s}{u[1 + 2\beta_s(1-q)] +
\beta_s(1 + s)^2}
\label{eq:hq} \\
q &= \frac{\frac12 \left[\rme^{-\hat{q}} \mbox{Erf}
\left(\frac{\alpha}{2\sqrt{\hat{u}}}
\right) - \rme^{\hat{q}}\mbox{Erf} \left(\frac{\alpha
-2\hat{s}}{2\sqrt{\hat{u}}}
\right) \right]-\sinh(\hat{q})}{\cosh(\hat{q}) +
\frac12 \left[\rme^{-\hat{q}} \mbox{Erf}
\left(\frac{\alpha}{2\sqrt{\hat{u}}}
\right) + \rme^{\hat{q}}\mbox{Erf} \left(\frac{\alpha
-2\hat{s}}{2\sqrt{\hat{u}}}
\right) \right]}
\label{eq:q} \\
s &= \frac{\frac1{\sqrt{\pi \hat{u}}}\left[\rme^{\hat{q}-[\frac{\alpha
-2\hat{s}}{2\sqrt{\hat{u}}}]^2}\right]}{\cosh(\hat{q}) + \frac12
\left[\rme^{-\hat{q}}\mbox{Erf} \left(\frac{\alpha }{2\sqrt{\hat{u}}}
\right) +
\rme^{\hat{q}}\mbox{Erf} \left(\frac{\alpha
-2\hat{s}}{2\sqrt{\hat{u}}} \right) \right]}
\label{eq:s} \\
u &= \frac{\frac1{4\sqrt{\pi \hat{u}^3}}\left[
\alpha\rme^{-\hat{q}-[\frac{\alpha}{2\sqrt{\hat{u}}}]^2} + (\alpha
-2 \hat{s})\rme^{\hat{q}-[\frac{\alpha
-2\hat{s}}{2\sqrt{\hat{u}}}]^2}\right]}{\cosh(\hat{q}) + \frac12
\left[\rme^{-\hat{q}}\mbox{Erf} \left(\frac{\alpha}{2\sqrt{\hat{u}}}
\right) + \rme^{\hat{q}} \mbox{Erf} \left(\frac{\alpha
-2\hat{s}}{2\sqrt{\hat{u}}} \right) \right]}
\label{eq:u}
\end{align}
\end{subequations}
\begin{figure}[htbp]
\begin{center}
\includegraphics[width=10cm]{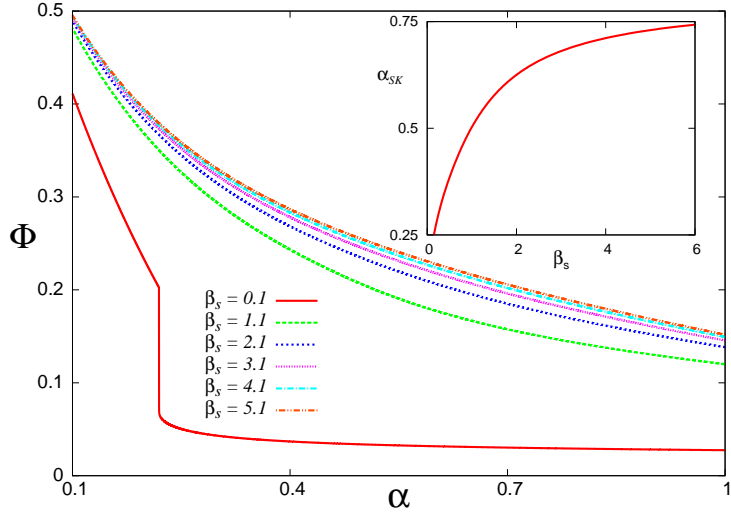}
\end{center}
\caption{\footnotesize
The logarithm of
the number of solution
$\Phi$
as a function of
$\alpha$ for several values of
$\beta_{s}$.
The inset is $\beta_{s}$-dependence of 
$\alpha_{\rm SK}$ at which $\Phi$ takes 
the same value as that of the SK model $\Phi_{\rm SK}=0.19923$\cite{TE}.}
\label{fig:fg1}
\end{figure}
We solve the equations (\ref{eq:hs})-(\ref{eq:u}) 
numerically for given values
of the parameters $(\beta_{s},\alpha)$. In Fig. \ref{fig:fg1}, we
plot the logarithm of the number, namely, 
the function $\Phi=\log \langle g_{0} \rangle/K$ at the saddle point
$(\hat{s}_{*}, \hat{u}_{*}, \hat{q}_{*},q_{*}, s_{*},u_{*})$. From
this figure, we find that the number of metastable solutions for the
zero-temperature dynamics (\ref{eq:zero_T_dynamics}) is larger than
that of the SK model ($\Phi_{\rm SK} \simeq 0.19923$) for large
numbers of users and small values of the Gaussian noise. As the
number of users $K=N/\alpha$ decreases ($\alpha$ increases), the
saddle point surface $\Phi$ tends to zero so the number of local
minimum solutions rapidly decreases. 
However, it remains finite for finite $K$. 

From Fig. \ref{fig:fg1}, we also see that the number of locally
minimum solutions decreases as the parameter $\beta_{s}$ decreases.
At a first glance, this seems to be rather counter-intuitive because
$\beta_{s}$ is the inverse of the variance of the Gaussian noise
which is defined by (\ref{eq:partition}). However, it might be
possible for us to show that this fact can be naturally understood.
We should notice that the Hamiltonian (\ref{eq:hamilton}) can be
rewritten as
\begin{eqnarray}
H(\bs) & = & \frac{1}{2} \sum_{ij} s_{i} J_{ij} s_{j} - \sum_{i}
h_{i} s_{i} \label{eq:Hamilton2}
\end{eqnarray}
where
$h_{i}$ is defined by
\begin{eqnarray}
h_{i} & = & \frac{1}{K} \sum_{\mu=1}^{N} \sum_{k=1}^{K} b_{k}^{\mu}
b_{i}^{\mu} s_{k}^{0} + \frac{\beta_{s}^{-1/2}}{\sqrt{K}}
\sum_{\mu=1}^{N} \eta^{\mu} b_{i}^{\mu}. \label{eq:random_field}
\end{eqnarray}
and where $\eta^{\mu}$ is a Gaussian variable with zero mean and
unit variance. The random field $h_{i}$ appearing in the above
Hamiltonian (\ref{eq:Hamilton2}) has zero mean and the variance
$\overline{h_{i}^{2}}=\alpha (1+\beta_{s}^{-1})$. Obviously, if
$\beta_{s} \ll 1$, the second term of (\ref{eq:Hamilton2}) becomes
dominant. Therefore, the best way to minimize the Hamiltonian
$H(\bs)$ is to minimize the second term of the Hamiltonian
(\ref{eq:Hamilton2}). In other words, making each bit $s_{i}$ in the
same direction as the random field $h_{i}$ is the best possible
strategy to minimize the Hamiltonian. As the result, the
frustration, which mainly comes from the first term of the
Hamiltonian, is weakened. This is a reason why the number of locally
stable solutions decreases as the parameter $\beta_{s}$ decreases.

We also should notice that the anomalous discontinuity of the
$\Phi$-$\alpha$ curve for $\beta_{s}=0.1$. These demodulation
problems as an annealed system 
have two locally stable solutions for $\Phi$ with large and
small overlap at the fixed point of the dynamics
(\ref{eq:zero_T_dynamics}). 
This result seems to imply that 
there exists a close relationship 
between the discontinuity of 
the $\Phi$-$\alpha$ curve and the spinodal 
observed in the bit-error rate \cite{Tanaka1,Tanaka2}. 
However, 
from the pioneering study by Tanaka \cite{Tanaka1}, 
one naturally expects the spinodal 
more likely to be observed when 
inter-user interference effects 
are more significant, namely, 
when the parameter $\beta_{s}$ is large. 
This counter-intuitive result might be 
caused due to our rough evaluation of 
the $\Phi$-$\alpha$ curve by annealed calculation. 
In fact, as we shall see later, 
the discontinuity 
disappears when we treat the problem as a quenched system. 
\section{Distribution of the local minimum energies}
Our next problem is to calculate the distribution of the energies of
these local minimum states. 
The local energy $-\epsilon_0$ is given by
\begin{eqnarray}
-\epsilon_0 = -\frac12 \sum_{\mu = 1}^N \left(y^\mu -
\frac{1}{\sqrt{K}} \sum_i b_i^\mu s_i \right)^2.
\end{eqnarray}
Then, the distribution of local energies is given by
\begin{multline}
\mathcal{N}(\epsilon_0) = \Bra \sum_{\bs} \prod_i\left[
\int_0^\infty \frac{\rmd \lambda_i}{g_0(\{b_i^\mu,z^\mu\})}
\delta\left(f_i - \sum_{j \neq i} J_{ij} s_j - \lambda_i \sigma_i
\right) \right] \right. \\
\times \left. \delta 
\left[\epsilon_0 -\frac12 \sum_{\mu = 1}^N
\left(y^\mu - \frac{1}{\sqrt{K}} \sum_i b_i^\mu s_i \right)^2
\right]
\Ket_{\{b_i^\mu,y^\mu\}}. \nonumber
\end{multline}
This is rather challenging to calculate directly, so we assume
rather that $g_0(\{b_i^\mu, y^\mu\})$ is self-averaging (following
Tanaka-Edwards \cite{TE,TE2}), or at least a slowly varying function of the
disorder and make the annealed approximation for this variable. We
then look at $\mathcal{P}(\epsilon_0) = \bra g_0 \ket
\mathcal{N}(\epsilon_0)$.
\begin{multline}
\mathcal{P}(\epsilon_0) = \Bra \sum_{\bs} \prod_i\left[
\int_0^\infty \rmd \lambda_i \delta\left(f_i - \sum_{j \neq i}
J_{ij} s_j - \lambda_i \sigma_i
\right) \right]\right. \\
\left. \times \delta \left[ \epsilon_0 -\frac12 \sum_{\mu = 1}^N
\left(y^\mu - \frac{1}{\sqrt{K}} \sum_i b_i^\mu s_i \right)^2
\right] \Ket_{\{b_i^\mu,y^\mu\}}\nonumber
\end{multline}
Now we can write the delta function constraining the energy in
Fourier representation as
\begin{multline}
\delta \left[ \epsilon_0 -\frac12 \sum_{\mu = 1}^N \left(y^\mu -
\frac{1}{\sqrt{K}} \sum_i b_i^\mu s_i \right)^2 \right] = \int
\frac{\rmd \hat{\epsilon}_0}{2\pi} 
{\exp}\left[
\rmi \hat{\epsilon}_0
\epsilon_0 - \frac{\rmi \hat{\epsilon}_0}{2} \sum_{\mu = 1}^N
\left(y^\mu - \frac{1}{\sqrt{K}} \sum_i b_i^\mu s_i \right)^2
\right]. 
\end{multline}
Then the calculation proceeds much as the previous one did, leading
to a saddle point surface given by:
\begin{multline}
\bra g_0 \ket = \left(\frac{\beta_s}{2\pi}
\right)^\frac{N}{2}\int \frac{\rmd t \rmd \hat{t}}{2\pi/K}
\frac{\rmd u \rmd \hat{u}}{2\pi/K} \frac{\rmd w \rmd
\hat{w}}{2\pi/K} \frac{\rmd q \rmd \hat{q}}{2\pi/K} \frac{\rmd
\hat{\epsilon}_0}{2\pi} \rme^{\rmi K
(\hat{t}t + \hat{u} u + \hat{w} w + \hat{q}q) + \rmi \hat{\epsilon}_0
\epsilon_0}\\
\times
\exp\Bigg[\alpha K \log \int \rmd y \frac{\rmd v \rmd
\hat{v}}{2\pi} \frac{\rmd v^0 \rmd \hat{v}^0}{2\pi}
\rme^{-\frac{\beta_s}{2} \left(y - v^0 \right)^2 + \rmi (\hat{v}^0
v^0 + \hat{v} v)- \frac12 \hat{v}^2 -\frac12 (\hat{v}^0)^2 -q\hat{v}
\hat{v}^0 -\frac{u}{2} (v - y)^2} \nonumber\\
\rme^{- t\hat{v}(v - y) -w \hat{v}^0 (v -
y) - \frac{\rmi \hat{\epsilon}_0}{2}(y - v)^2} \Bigg]\nonumber \\
  \times \exp\Bigg[K \log \frac12 \sum_{s,s^0}\int_0^\infty \rmd \lambda
\int_{-\infty}^{\infty} \frac{\rmd \hat{\lambda}}{2\pi} \rme^{\rmi
\hat{\lambda} (\alpha - \lambda) - \rmi (\hat{q} s s^0 + \hat{u}
\hat{\lambda}^2 + \hat{t} \hat{\lambda} + \hat{w} \hat{\lambda} s
s^0)}\Bigg]
\end{multline}
From here we rotate $\hat{\epsilon}_0 \to -\rmi \hat{\epsilon}_0$
and then rescale $\epsilon_0 \to \epsilon_0/K$. 
Hence, from now on,  we call the $\epsilon_{0}$ as 
``normalized-energy". It is relatively
straightforward to read off the final result:
\begin{multline}
\mathcal{P}(\epsilon_0) =\int \frac{\rmd t \rmd \hat{t}}{2\pi/K}
\frac{\rmd u \rmd \hat{u}}{2\pi/K} \frac{\rmd w \rmd
\hat{w}}{2\pi/K} \frac{\rmd q \rmd \hat{q}}{2\pi/K} \frac{\rmd
\hat{\epsilon}_0}{2\pi} \rme^{ K \Psi(
t,u,w,q,\hat{t},\hat{u},\hat{w},\hat{q},\epsilon_0,\hat{\epsilon}_0)} \\
\Psi = \hat{t}t + \hat{u} u + \hat{w} w + \hat{q}q +
\hat{\epsilon}_0 \epsilon_0 -\frac{\alpha}{2} \log \left\{ (u +
\hat{\epsilon}_0)[1 + 2\beta_s(1-q)] + \beta_s(1 + t
-w)^2\right\}\nonumber
\\ + \log \left\{\cosh(\hat{q}) + \frac12
\left[\rme^{-\hat{q}}\mbox{Erf} \left(\frac{\alpha -\hat{t} -
\hat{w}}{2\sqrt{\hat{u}}} \right) + \rme^{\hat{q}}\mbox{Erf}
\left(\frac{\alpha -\hat{t} + \hat{w}}{2\sqrt{\hat{u}}} \right)
\right]\right\} + \frac{\alpha}{2} \log \beta_s
\end{multline}
From the following equations we see that the saddle point equations
will be very similar to those we obtained for the calculation of the
number of solutions ${\exp}[N\Phi]$ only now with $u \to u +
\hat{\epsilon}_0$. Varying with respect to 
$\hat{\epsilon}_0$ leads to the
conclusion that we must have $\epsilon_0 = \hat{u}$. 
We can make the same substitutions as before with $s$ and $\hat{s}$
to find
\begin{multline}
\Psi = \hat{s}s + \hat{u} u + \hat{q}q + \hat{\epsilon}_0
\epsilon_0 -\frac{\alpha}{2} \log \left\{ (u + \hat{\epsilon}_0)[1 +
2\beta_s(1-q)] + \beta_s(1 + s)^2\right\}\nonumber \\
+ \log \left\{\cosh(\hat{q}) + \frac12
\left[\rme^{-\hat{q}}\mbox{Erf} \left(\frac{\alpha}{2\sqrt{\hat{u}}}
\right) + \rme^{\hat{q}}\mbox{Erf} \left(\frac{\alpha
-2\hat{s}}{2\sqrt{\hat{u}}} \right) \right]\right\} +
\frac{\alpha}{2} \log \beta_s.
\end{multline}
Thus, the saddle point equations are given by
\begin{subequations}
\begin{align}
\epsilon_0 &= \frac12\frac{\alpha[1 + 2\beta_s(1-q)] }{(u +
\hat{\epsilon}_0)[1 + 2\beta_s(1-q)] +
\beta_s(1 + s)^2}\label{eq:epsilon_SPe} \\
\hat{s} &= \frac{\alpha \beta_s (1 +s )}{(u +
\hat{\epsilon}_0)[1 + 2\beta_s(1-q)] +
\beta_s(1 + s)^2} \label{eq:hats_SPe} \\
\hat{u} &= \frac12\frac{\alpha[1 + 2\beta_s(1-q)] }{(u +
\hat{\epsilon}_0)[1 + 2\beta_s(1-q)] +
\beta_s(1 + s)^2} = \epsilon_0 \\
\hat{q} &= \frac{-\alpha \beta_s(u + \hat{\epsilon}_0)}{(u +
\hat{\epsilon}_0)[1 + 2\beta_s(1-q)] +
\beta_s(1 + s)^2}\\
q &= \frac{\frac12 \left[\rme^{-\hat{q}} \mbox{Erf}
\left(\frac{\alpha}{2\sqrt{\hat{u}}}
\right) - \rme^{\hat{q}}\mbox{Erf} \left(\frac{\alpha
-2\hat{s}}{2\sqrt{\hat{u}}} \right)
\right]-\sinh(\hat{q})}{\cosh(\hat{q}) + \frac12
\left[\rme^{-\hat{q}} \mbox{Erf}
\left(\frac{\alpha}{2\sqrt{\hat{u}}} \right) +
\rme^{\hat{q}}\mbox{Erf} \left(\frac{\alpha
-2\hat{s}}{2\sqrt{\hat{u}}}
\right) \right]} \label{eq:q_SPe} \\
s & =\frac{\frac1{\sqrt{\pi \hat{u}}}\left[\rme^{\hat{q}-[\frac{\alpha
-2\hat{s}}{2\sqrt{\hat{u}}}]^2}\right]}{\cosh(\hat{q}) + \frac12
\left[\rme^{-\hat{q}}\mbox{Erf} \left(\frac{\alpha
}{2\sqrt{\hat{u}}} \right) + \rme^{\hat{q}}\mbox{Erf}
\left(\frac{\alpha
-2\hat{s}}{2\sqrt{\hat{u}}} \right) \right]} \label{eq:s_SPe}\\
u &= \frac{\frac1{4\sqrt{\pi \hat{u}^3}}\left[
\alpha\rme^{-\hat{q}-[\frac{\alpha}{2\sqrt{\hat{u}}}]^2} + (\alpha
-2 \hat{s})\rme^{\hat{q}-[\frac{\alpha
-2\hat{s}}{2\sqrt{\hat{u}}}]^2}\right]}{\cosh(\hat{q}) + \frac12
\left[\rme^{-\hat{q}}\mbox{Erf} \left(\frac{\alpha}{2\sqrt{\hat{u}}}
\right) + \rme^{\hat{q}} \mbox{Erf} \left(\frac{\alpha
-2\hat{s}}{2\sqrt{\hat{u}}} \right) \right]} \label{eq:u_SPe}
\end{align}
\end{subequations}
In order to solve these saddle point equations, we evaluate the
fixed points for a given value of $\hat{\epsilon}_0$, then calculate
the value of the normalized-energy 
$\epsilon_0$, and finally read off the value of
$n(\epsilon_0)$ for 
the scaling form of the distribution 
${\cal N}(\epsilon_{0})
\simeq {\exp}[K(\Psi-\Phi)] =[n(\epsilon_{0})]^{K}$, $\log
n(\epsilon_{0}) \equiv \Psi-\Phi$ 
and then change $\hat{\epsilon}_0$ and repeat. This
is more straightforward than trying to find the correct
$\hat{\epsilon}_0$ for a given $\epsilon_0$. 
Thus,
for the solution of the saddle point equations
(\ref{eq:epsilon_SPe})-(\ref{eq:u_SPe}),
the distribution
we seek to obtain is
given by
\begin{multline}
\log n(\epsilon_{0}) =
\hat{\epsilon}_{0} \epsilon_{0}
-\frac{\alpha}{2}\log
\left\{
(u +\hat{\epsilon}_{0})
[1+2\beta_{s}(1-q)] +
\beta_{s}(1+s)^{2} \right\} \\
+ \frac{\alpha}{2}\log
\left\{
u [1+2\beta_{s}(1-q)] +
\beta_{s}(1+s)^{2} \right\}.
\end{multline}
where the variables on the right hand side obviously take their saddle point
values.
\begin{figure}[ht]
\begin{center}
\includegraphics[width=7.7cm]{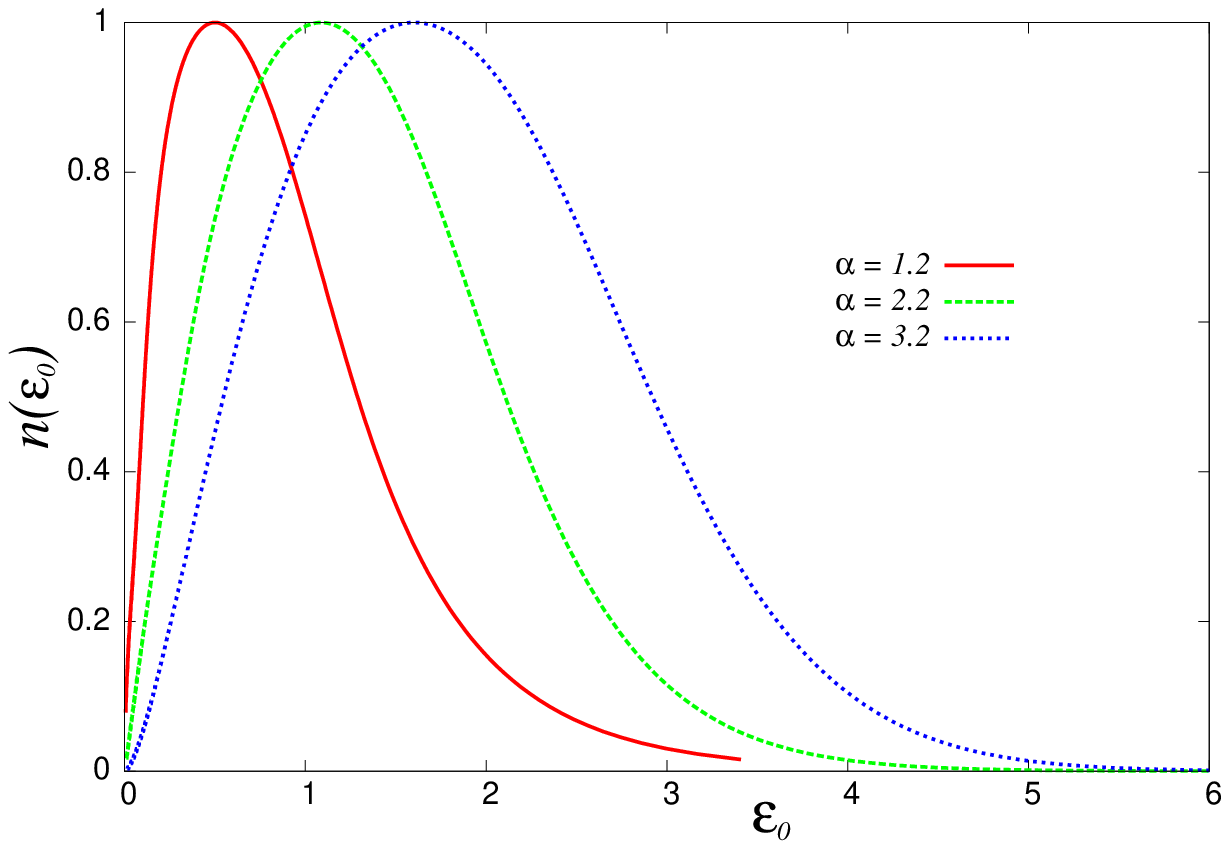}
\includegraphics[width=7.7cm]{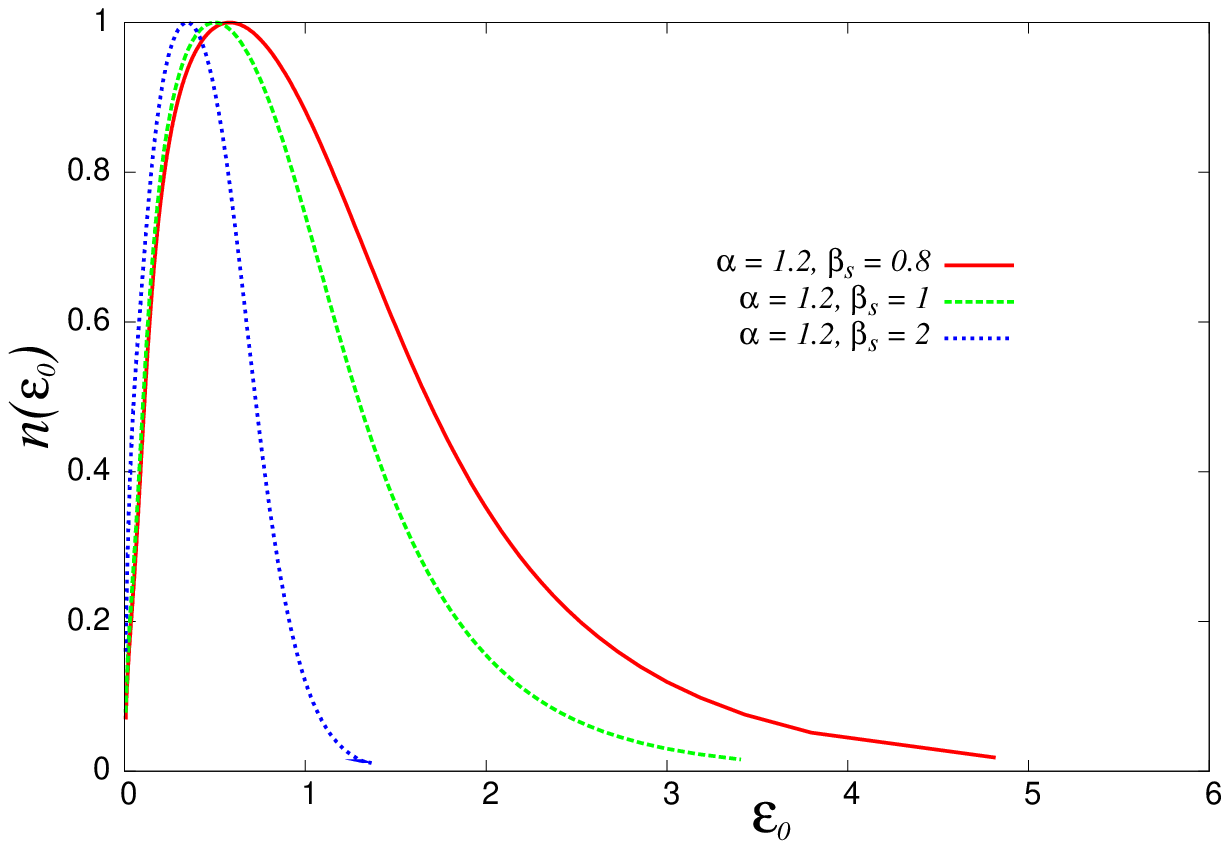}
\end{center}
\caption{\footnotesize The distribution of the minimum 
normalized-energy states. In the left panel, the variance 
of the Gaussian noise
$\beta_{s}$ is fixed to $1.0$ and we vary $\alpha$ as $1.2,2.2$ and
$3.2$. The right panel is obtained by setting $\alpha=1.2$ and
varying $\beta_{s}$ as $2,1$ and $0.8$. } \label{fig:fg2}
\end{figure}
In Fig. \ref{fig:fg2}, we plot the distribution $n(\epsilon_{0})$.
From the left panel of this figure, we find that the probability for
deep normalized-energy state $\epsilon_{0} \gg 1$ is 
almost zero for all values of the number
of users $\alpha$, however, relatively higher normalized-energy states appear
much more frequently as the number of users increases ($\alpha$
decreases). On the other hand, the right panel tells us that the
deep energy states frequently appear in the case of large variance
$\beta_{s}^{-1}$ of the Gaussian. The reason why the deep energy
states frequently appear as the variance of the Gaussian noise
$\beta_{s}^{-1}$ increases is the same reason as we explained for
the $\beta_{s}$-dependence of the locally minimum solutions. For
large value of $\beta_{s}^{-1}$, some of the bits $b_{i}$ take their
direction so as to take the same sign as that of the local field
$h_{i}$ whose strength is estimated as $\alpha (1+\beta_{s}^{-1})$.
As a result, such a bit becomes free from frustration effects and
hence, the deep energy state appears more frequently. However, if
$\beta_{s}^{-1}$ decreases, the effect of the random field term
$\sum_{i}h_{i}s_{i}$ is weakened and the frustration has more effect
on some of the bits. Therefore, the $\beta_{s}^{-1}$-dependence of
the distribution shown in the right panel of Fig. \ref{fig:fg2} is
clearly understood, although we should note that of course this does
not necessarily mean that a better decoding result is possible at
higher noise levels, just that lower energy solutions are more
accessible.
\section{Beyond the annealed approach}
In section 3, 
we considered the quantity $g_{0} \sim 
{\rm e}^{K \Phi (\{J,f\})}$ 
for a given realization of 
disorder $\{J,f\}(=\{b_{i}^{\mu},y^{\mu}\})$ and 
assumed that 
the value $g_{0}$ is identical to 
the average $\langle g_{0} \rangle$ 
in thermodynamic limit. 
However, 
for a specific choice of the disorder 
$\{J,f\}$, 
the $g_{0}$ might take an extremely 
large value of exponential order of $K$ 
and it is difficult for us to 
confirm that such an extreme value coincides with 
the average $\langle g_{0} \rangle$. 
For this reason, 
we should take the quantity 
$\log g_{0}$ instead of $g_{0}$ and 
assume that $\log g_{0} = 
K \Phi (\{J,f\})$ for a given $\{J,f\}$ 
should be equal to 
the average $\langle \log g_{0} \rangle$ 
in thermodynamic limit. 
Then, the $\log g_{0}$ does not take extremely large values 
in comparison with the typical value of $\log g_{0}$ 
even if we choose a specific choice of 
the disorder $\{J,f\}$.  

However, this will,
unfortunately, require more technology and a more involved
calculation as we will have to introduce replica theory
\cite{Nishimori}. In some ways the previous sections can be viewed
as an introductory calculation. Now,
\begin{eqnarray}
g_0(\{ b_{i}^{\mu}, y^{\mu} \}) = \sum_{\bs} \prod_i \left[ \int_0^\infty
\rmd
\lambda_{i} \delta\left(f_i - \sum_{j \neq i} J_{ij} s_j - \lambda_i
\sigma_{i} \right) \right]
\end{eqnarray}
and we wish to calculate
\begin{eqnarray}
\bra \log g_0(\{b_{i}^{\mu}, y^{\mu} \}) 
\ket_{\{b_i^\mu, y^{\mu} \}} =
\lim_{n \to 0} \frac1n (\bra g_0^n(\{b_{i}^{\mu}, y^{\mu} \})
\ket_{\{b_{i}^{\mu}, y^{\mu} \}} - 1).
\end{eqnarray}
using the powerful replica approach. Then, following the usual
algebra (similar in many respects to the earlier sections), we have
\begin{eqnarray}
\bra g_0^n \ket = \int \mathcal{D}(m,w,r,q,u) \rme^{K
\Phi(\{m,w,q,u,\hat{m},\hat{w},\hat{q},\hat{u}\})}
\label{eq:g0_n}
\end{eqnarray}
with
\begin{eqnarray}
\Phi & = & \rmi \sum_{\alpha > 0} 
(\hat{m}^\alpha m^\alpha +
\hat{w}^\alpha w^\alpha) + \rmi \sum_{\alpha < \beta}
\hat{q}^{\alpha \beta} q^{\alpha \beta} + \rmi \sum_{\alpha \leq
\beta} \hat{u}^{\alpha \beta} u^{\alpha \beta} + \rmi \sum_{\alpha
\beta} \hat{r}^{\alpha \beta} r^{\alpha \beta} + \frac{\alpha}{2}
\log \frac{\beta_s}{2\pi} \nonumber \\
\mbox{} & + & \log \frac12 \sum_{s^0,s^1,\ldots,s^n} \prod_{\alpha}
\left[\int
\frac{\rmd \lambda^\alpha \rmd \hat{\lambda}^\alpha}{2\pi}
\right] 
\exp
{\Biggr [} \rmi \sum_{\alpha} \hat{\lambda}^\alpha (\alpha -
\lambda^\alpha) - \rmi s^0 \sum_\alpha (\hat{m}^\alpha s^\alpha+
\hat{w}^\alpha s^\alpha \hat{\lambda}^\alpha) \nonumber \\
\mbox{} & - & \rmi \sum_{\alpha <
\beta} s^\alpha s^\beta \hat{q}^{\alpha \beta} - \rmi \sum_{\alpha
\leq \beta} \hat{u}^{\alpha \beta} s^\alpha s^\beta
\hat{\lambda}^\alpha \hat{\lambda}^\beta - \rmi \sum_{\alpha \beta}
\hat{r}^{\alpha \beta} s_\alpha  s_\beta
\hat{\lambda}^\beta
{\Biggr ]} \nonumber \\
\mbox{} & + & \alpha \log \int \left[\rmd y
\prod_{\alpha = 0}^n \frac{\rmd v^\alpha \rmd \hat{v}^\alpha}{2\pi}
\right]
\exp\left[-\frac{\beta_s}{2} \left(y - v^0 \right)^2 + \rmi
\sum_{\alpha = 0}^n \hat{v}^\alpha v^\alpha
\right] \nonumber \\
\mbox{} & \times & 
\exp 
{\Biggr [}- \frac{1}{2} [(\hat{v}^0)^2+ \sum_{\alpha \beta > 0}
[q^{\alpha \beta} \hat{v}^\alpha \hat{v}^\beta + 2 r^{\alpha \beta}
\hat{v}^\alpha (v^\beta - y) + u^{\alpha \beta} (v^\alpha -
y)(v^\beta - y)] \nonumber \\
\mbox{} & + & 2\hat{v}^0 \sum_{\alpha > 0} [m^\alpha
\hat{v}^\alpha + w^\alpha(v^\alpha - y)]]
{\Biggr ]}
\label{eq:quench_surface}
\end{eqnarray}
where we defined
$m^{\alpha},
w^{\alpha},
q^{\alpha \beta},
u^{\alpha \beta},
r^{\alpha \beta}$ as
\begin{subequations}
\begin{align}
m^{\alpha} &=
\frac{1}{K}\sum_{k}
s_{k}^{0}
s_{k}^{\alpha}
\label{eq:hm2_SP} \\
w^{\alpha} &=
\frac{1}{K}\sum_{k}s_{k}^{0}
s_{k}^{\alpha}
\hat{\lambda}_{k}^{\alpha} \\
q^{\alpha \beta} &=
\frac{1}{k}
\sum_{k}s_{k}^{\alpha}
s_{k}^{\beta} \\
u^{\alpha \beta} &=
\frac{1}{K}
\sum_{k}
s_{k}^{\alpha}
s_{k}^{\beta}
\hat{\lambda}_{k}^{\alpha}
\hat{\lambda}_{k}^{\beta} \\
r^{\alpha \beta} &=
\frac{1}{K}
\sum_{k}
s_{k}^{\alpha}
s_{k}^{\beta}
\hat{\lambda}_{k}^{\beta}
\label{eq:r2_SP}
\end{align}
\end{subequations}
and their conjugates : $\hat{m}^{\alpha}, \hat{w}^{\alpha},
\hat{q}^{\alpha \beta}, \hat{u}^{\alpha \beta}, \hat{r}^{\alpha
\beta}$ by introducing the definitions
(\ref{eq:hm2_SP})-(\ref{eq:r2_SP}) via integral representation of
delta function as e.g.
\begin{eqnarray}
\int \prod_{1\leq\alpha \leq n} \left[\frac{\rmd m^\alpha
\rmd \hat{m}^\alpha}{2\pi/K}\right]
\exp \left[
\rmi K\sum_\alpha
\left[ \hat{m}^\alpha 
\left(
m^\alpha - \frac{1}{K}
\sum_k s_k^0 s_k^\alpha
\right)
\right]
\right] = 1.
\end{eqnarray}
We also introduced the shorthand
$\mathcal{D}(m,w,r,q,u)$
to indicate integral over these
saddle point variables.

Then, the replica symmetric ansatz simplifies the saddle point
defining $\langle g_{0}^{n} \rangle$. After a relatively involved
calculation, we obtain
\begin{multline}
\frac{1}{n} \Phi_{RS} = \hat{m}m+ \hat{w} w + \frac{1}{2}
\hat{q} q + \frac12 \hat{u}_d u_d - \frac{1}{2}
\hat{u}u - \hat{r}_d r_d + \hat{r}r\\
-\frac{\hat{q}}{2}  + \frac12 \sum_{s^0} \int Dz_1 Dz_2 Dz_3
\log \Bigg\{\frac12 \sum_s \rme^{-s^0 s \hat{m} + z_1 \sqrt{\hat{q}
- \hat{r}} s + z_3 \sqrt{\hat{r}} s} \Bigg[1  \\
+ \mbox{Erf}\Bigg(\frac{ \alpha + \hat{r}_d  - \hat{r} - s s^0
\hat{w} + z_2 \sqrt{\hat{u} - \hat{r}} s + z_3 \sqrt{\hat{r}}
s}{\sqrt{2(\hat{u}_d - \hat{u})}}
\Bigg) \Bigg]\Bigg\}\\
+ \alpha \Bigg\{ - \frac12 \log[(1 + r_d - r)^2 +
(u_d - u)(1 - q)] \\
+ \frac{2(w-r)(1 + r_d - r)
-u(1-q) -(1 + \beta_s^{-1} + q - 2m)(u_d - u) }{2\{(1 + r_d - r)^2 +
(u_d - u)(1 - q)\}} \Bigg\}
\label{eq:quench_surface_rs}
\end{multline}
The details of the derivation is explained in Appendix B.

We next derive the saddle point equations. By introducing two
short-hand measures:
\begin{multline}
\bra \ldots \ket_1 \equiv \frac12 \sum_{s^0} \int Dz_1 Dz_2 Dz_3
\frac{\sum_s  \ldots \rme^{-s^0 s \hat{m} + z_1 \sqrt{\hat{q} -
\hat{r}} s + z_3 \sqrt{\hat{r}} s} \Bigg[1 + \mbox{Erf}\Bigg(\frac{
\alpha + \hat{r}_d - \hat{r} - s s^0 \hat{w} + z_2 \sqrt{\hat{u} -
\hat{r}} s + z_3 \sqrt{\hat{r}} s}{\sqrt{2(\hat{u}_d - \hat{u})}}
\Bigg) \Bigg] }{\sum_s \rme^{-s^0 s \hat{m} + z_1 \sqrt{\hat{q} -
\hat{r}} s + z_3 \sqrt{\hat{r}} s} \Bigg[1 + \mbox{Erf}\Bigg(\frac{
\alpha + \hat{r}_d - \hat{r} - s s^0 \hat{w} + z_2 \sqrt{\hat{u} -
\hat{r}} s + z_3 \sqrt{\hat{r}} s}{\sqrt{2(\hat{u}_d - \hat{u})}}
\Bigg) \Bigg]} \nonumber \\
\bra \ldots \ket_2 \equiv \frac1{\sqrt{\pi}} \sum_{s^0} \int Dz_1 Dz_2
Dz_3 \frac{\sum_s  \ldots \rme^{-s^0 s \hat{m} + z_1 \sqrt{\hat{q} -
\hat{r}} s + z_3 \sqrt{\hat{r}} s -[\frac{ \alpha + \hat{r}_d -
\hat{r} - s s^0 \hat{w} + z_2 \sqrt{\hat{u} - \hat{r}} s + z_3
\sqrt{\hat{r}} s}{\sqrt{2(\hat{u}_d - \hat{u})}}]^2}}{\sum_s
\rme^{-s^0 s \hat{m} + z_1 \sqrt{\hat{q} - \hat{r}} s + z_3
\sqrt{\hat{r}} s} \Bigg[1 + \mbox{Erf}\Bigg(\frac{ \alpha + \hat{r}_d
- \hat{r} - s s^0 \hat{w} + z_2 \sqrt{\hat{u} - \hat{r}} s + z_3
\sqrt{\hat{r}} s}{\sqrt{2(\hat{u}_d - \hat{u})}} \Bigg)
\Bigg]}\nonumber
\end{multline}
and varying our conjugate order parameters,
we obtain
\begin{subequations}
\begin{align}
q &= 1 - \left\bra \frac{z_1 s}{\sqrt{\hat{q} - \hat{r}}} \right\ket_1
\label{eq:q_quench} \\
m &= \bra s s_0 \ket_1 \\
w &= \left\bra \frac{s s_0}{\sqrt{2(\hat{u}_d - \hat{u})}} 
\right\ket_2 \\
u &= \left\bra \frac{z_2 s}{\sqrt{2(\hat{u}_d - \hat{u})(\hat{u} -
\hat{r})}} \right\ket_2 + 2 
\left\bra \frac{\alpha + \hat{r}_d - \hat{r} - s s^0
\hat{w} + z_2 \sqrt{\hat{u} - \hat{r}}s + z_3
\sqrt{\hat{r}}s}{[2(\hat{u}_d - \hat{u})]^{\frac32}} 
\right\ket_2 \\
u_d &= \left\bra \frac{\alpha + \hat{r}_d - \hat{r} - s s^0 \hat{w} + z_2
\sqrt{\hat{u} - \hat{r}}s + z_3
\sqrt{\hat{r}}s}{[2(\hat{u}_d - \hat{u})]^{\frac32}}
\right\ket_2 \\
r & = \left\bra \frac{z_1 s}{2\sqrt{\hat{q}- \hat{r}}}  - \frac{z_3
s}{2\sqrt{\hat{r}}}
\right\ket_1 + 
\left\bra \frac{z_2 s}{2\sqrt{2(\hat{u}_d
-\hat{u})(\hat{u} - \hat{r})}} - \frac{z_3 s}{2\sqrt{2(\hat{u}_d -
\hat{u})\hat{r}}}  + \frac{1}{\sqrt{2(\hat{u}_d - \hat{u})}}
\right\ket_2 \\
r_d &= \left\bra \frac{1}{\sqrt{2(\hat{u}_d - \hat{u})}} 
\right\ket_2
\label{eq:rd_quench}
\end{align}
\end{subequations}
We introduce a further minor shorthand:
\begin{subequations}
\begin{align}
A &\equiv  (1 + r_d
-r)^2
+ (u_d - u)(1-q) \\
B &\equiv 2(w-r)(1 + r_d - r) -u(1-q) -(1 + \beta_s^{-1} + q -
2m)(u_d - u)
\end{align}
\end{subequations}
By varying the order parameters themselves, we have
\begin{subequations}
\begin{align}
\hat{m} &= \frac{-\alpha(u_d - u)}{A}\\
\label{eq:hat_m_quench}
\hat{w} &= \frac{-\alpha(1 + r_d -r)}{A}\\
\hat{q} &= \frac{-\alpha u}{A} - \frac{\alpha B(u_d - u)}{A^2}\\
\hat{u}_d  &= \frac{\alpha (2 + \beta_s^{-1} - 2m)}{A} + \frac{\alpha
B(1-q)}{A^2}\\
\hat{u} & = \frac{\alpha(1 + \beta_s^{-1} + q - 2m)}{A} + \frac{\alpha
B(1-q)}{A^2}\\
\hat{r}_d & = \frac{\alpha(w - 1 - r_d)}{A} - \frac{\alpha B(1 + r_d -
r)}{A^2} \\
\hat{r} &= \frac{\alpha(w -r)}{A} -\frac{\alpha B(1 + r_d - r)}{A^2}
\label{eq:hat_r_quench}
\end{align}
\end{subequations}
We solve these fourteen equations
(\ref{eq:q_quench})-(\ref{eq:rd_quench}) and
(\ref{eq:hat_m_quench})-(\ref{eq:hat_r_quench}) numerically to
obtain the average number of locally minimum states of the CDMA
multiuser MAP demodulator treated as a quenched system.
\begin{figure}[ht]
\begin{center}
\includegraphics[width=10cm]{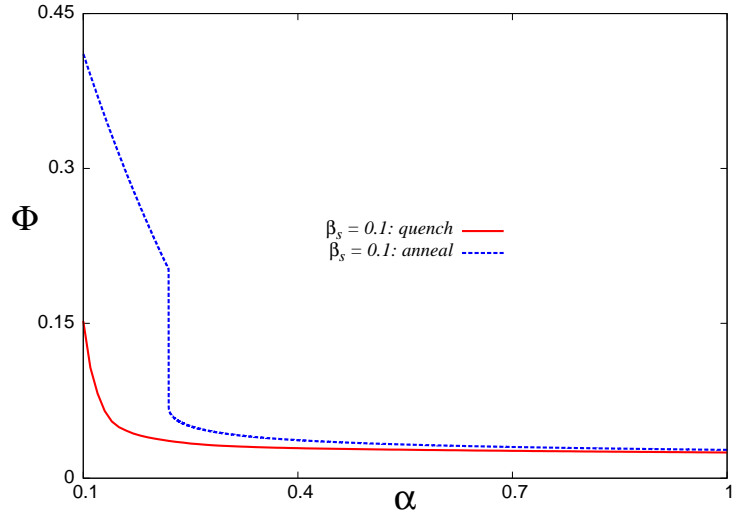}
\end{center}
\caption{\footnotesize
The average number of
locally minimum solutions
evaluated by a quenched calculation. The logarithm of the number,
$\Phi$, is plotted as a function of $\alpha$ for
$\beta_{s}=0.1$.} 
\label{fig:fg3}
\end{figure}
In Fig.\ref{fig:fg3}, we plot $\Phi$, that is, the logarithm of the
average number of locally minimum solutions evaluated by the
quenched calculation as a function of $\alpha$ for $\beta_{s}=0.1$.
From this figure, we find that the number of solutions decreases in
comparison with the results found in the annealed calculation. This
result can be confirmed by the following argument. Convexity of the
logarithm gives
\begin{eqnarray}
\langle \log g_{0} \rangle & \leq &
\log \langle g_{0} \rangle
\end{eqnarray}
and the logarithm of the average number of the MAP solutions
evaluated by the annealed approximation should be larger than the
result of the quenched calculation. Taking into account this fact,
The result shown in Fig. \ref{fig:fg3} is quite natural. 
It should be noted that the discontinuity 
observed in the annealed calculation 
disappears for the quenched evaluation. 
From these
results we conclude that within a much more precise treatments than
the annealed calculation 
in the sense that 
self-averaging quantity here is not 
$g_{0}$ but $\log g_{0}$, the average number of locally minimum
solutions of the CDMA multiuser MAP demodulator 
continuously decreases as $\alpha$ increases, however, it
is still of exponential order. 
The result we obtained here might provide useful
information about the computational complexity which could help in 
constructing sophisticated 
algorithms to 
obtain the solution for the CDMA multiuser MAP demodulator.
\section{Summary}
In this paper, we investigated the ground state properties of the
CDMA multiuser demodulator, in particular, the number of locally
minimum solutions of the zero-temperature dynamics of the MAP
demodulator by both annealed and replica symmetric calculations.
Moreover, we evaluated the distribution of the local minimum 
normalized-energy.
We found that the number of locally stable solutions of the MAP
demodulator is larger than that of the SK model for a large number
of users and small values of the deviation of the Gaussian noise. We
also found that when the number of the users $K$ decreases, the
saddle point surface also decreases. However, it never reaches 
zero for finite $K$ and as a result, the number of
solutions turns out to be exponential order. From these results, 
we might have
useful information when we attempt to construct an algorithm to
search for the ground state of the CDMA Hamiltonian
(\ref{eq:hamilton}). From the evaluation of the
distribution of the local minimum normalized-energies, we found that the
probability for deep normalized-energy state $\epsilon_{0} \gg 1$ is 
almost zero for all values of 
the number of the users $\alpha$, however, relatively
higher energy states appear much more frequently as the number of
the users increases ($\alpha$ decreases). The analysis also told us
that the deep energy states frequently appears for the case of large
variance $\beta_{s}^{-1}$ of the Gaussian. 

We hope that our analysis
here provides a useful guide for the engineers to construct the MAP
demodulator for CDMA systems.
\section*{Acknowledgments}
One of the authors (J.I.) was financially supported by 
Grant-in-Aid for Scientific Research on Priority Areas: 
{\it Statistical-Mechanical Approach to 
Probabilistic Information Processing (SMAPIP)} and 
{\it Deepening and Expansion of 
Statistical Mechanical Informatics (DEX-SMI)} of 
The Ministry of Education, Culture, 
Sports, Science and Technology (MEXT) 
No. 14084201 and No. 18079001, respectively. 
J.H. would
like to thank Hokkaido University for its hospitality, where most of
this work was completed. This paper has been prepared by the authors
as a commentary on the topic as at November 2006 and is not a
definitive analysis of the subject matter covered.  It does not
constitute advice or guidance in this area and should not be relied
upon as such.  The paper has not been prepared by the J.H. in his
capacity as an employee of Hymans Robertson LLP and the views
expressed do not represent those of Hymans Robertson LLP. 
Neither the authors nor Hymans Robertson LLP accept 
liability for any errors or omissions. 

We thank anonymous referees for 
a lot of constructive comments on 
the manuscript. 
\appendix
\section{Derivation of the average number of solutions
by annealed calculation}
In this appendix, we explain the details of the derivation of the
average number of locally minimum solutions from the definition
(\ref{eq:anneal_g0}). First of all, we introduce new variables
$v_{\mu}^{0}$ and $v_{\mu}$ by inserting the equations:
\begin{eqnarray}
1 = \prod_\mu \int \rmd v^0_\mu  \delta
\left[v^0_\mu - \frac{1}{\sqrt{K}}
\sum_{k = 1}^K b_k^\mu s_k^0
\right]
= \int \prod_\mu \left[\frac{\rmd v^0_\mu \rmd \hat{v}^0_\mu}{2\pi}
\right] \exp
\left[
\rmi \sum_\mu \hat{v}^0_\mu 
\left(v^0_\mu
-\frac{1}{\sqrt{K}} \sum_{k = 1}^K b_k^\mu s_k^0
\right)
\right]
\end{eqnarray}
and
\begin{eqnarray}
1 = \int \prod_\mu 
\left[\frac{\rmd v_\mu \rmd \hat{v}_\mu}{2\pi}
\right] \exp
\left[
\rmi \sum_\mu \hat{v}_\mu 
\left(
v_\mu -\frac{1}{\sqrt{K}}
\sum_{k = 1}^K b_k^\mu s_k
\right)
\right].
\end{eqnarray}
Then, equation (\ref{eq:anneal_g0}) is
rewritten as
\begin{multline}
\bra g_0 \ket = 2^{-K}\left(\frac{\beta_s}{2\pi}
\right)^\frac{N}{2} \sum_{\bs^0} \int \prod_{\mu = 1}^N \left[\rmd
y^\mu \frac{\rmd v_\mu \rmd \hat{v}_\mu}{2\pi} \frac{\rmd v^0_\mu
\rmd \hat{v}^0_\mu}{2\pi} \right]
\exp 
\left[
-\frac{\beta_s}{2} \sum_{\mu
= 1}^N \left(y^\mu -
v_\mu^0 \right)^2
\right] 2^{-NK} \\
\times 
\sum_{\bs, \bb^1,\ldots,\bb^N} \prod_i \int_0^\infty
\rmd \lambda_i \int_{-\rmi\infty}^{\rmi\infty} \frac{\rmd
\hat{\lambda}_i}{2\pi \rmi} \exp\left[\hat{\lambda}_i
\left\{\frac{1}{\sqrt{K}} \sum_\mu y^\mu b_i^\mu -
\frac{1}{\sqrt{K}}\sum_\mu b_i^\mu 
\left(
v_\mu - \frac{1}{\sqrt{K}}
b_i^\mu s_i
\right) - \lambda_i s_i \right\} \right]\nonumber\\
\times \exp
\left[
\rmi \sum_\mu \hat{v}_\mu 
\left(
v_\mu -\frac{1}{\sqrt{K}} \sum_{k =
1}^K b_k^\mu s_k
\right)
+\rmi \sum_\mu \hat{v}^0_\mu 
\left(
v^0_\mu
-\frac{1}{\sqrt{K}} \sum_{k = 1}^K b_k^\mu s_k^0
\right)
\right]
\end{multline}
where we used
$(b_{i}^{\mu})^{2}=1$.
We next focus on the average in the last line 
in the limit of $K \to \infty$:
\begin{multline}
2^{-NK} \sum_{\bb^1,\ldots,\bb^N} 
\exp
\left[-\frac{\rmi}{\sqrt{K}}
\sum_\mu \sum_{k = 1}^K b_k^\mu(\hat{v}_\mu s_k+ \hat{v}^0_\mu s_k^0
+ \hat{\lambda}_k v^\mu - \hat{\lambda}_k y^\mu )
\right] \\
= \prod_{\mu k} \exp
\left[\log \cosh
\left[-\frac{\rmi}{\sqrt{K}}(\hat{v}_\mu
s_k+ \hat{v}^0_\mu s_k^0 +\hat{\lambda}_k v^\mu - \hat{\lambda}_k
y^\mu )
\right]
\right] \\
\simeq \exp
\left[-\frac{1}{2K} \sum_{\mu k}(\hat{v}_\mu s_k+ \hat{v}^0_\mu
s_k^0 +\hat{\lambda}_k v^\mu - \hat{\lambda}_k y^\mu )^2
\right]
\end{multline}
and this reads
\begin{eqnarray}
\bra g_0 \ket & = & 2^{-K}\left(\frac{\beta_s}{2\pi}
\right)^\frac{N}{2} \int \prod_{\mu = 1}^N \left[\rmd y^\mu
\frac{\rmd v_\mu \rmd \hat{v}_\mu}{2\pi} \frac{\rmd v^0_\mu \rmd
\hat{v}^0_\mu}{2\pi} \right] \nonumber \\
\mbox{} & \times & 
\exp \left[-\frac{\beta_s}{2} \sum_{\mu =
1}^N \left(y^\mu -
v_\mu^0 \right)^2
+ \rmi \sum_\mu (\hat{v}^0_\mu v^0_\mu + \hat{v}_\mu v_\mu)
\right] \nonumber \\
\mbox{} & \times & \sum_{\bs,\bs^0}  
\prod_i\left[\int_0^\infty \rmd \lambda_i
\int_{-\infty}^{\infty} \frac{\rmd \hat{\lambda}_i}{2\pi}\right]
\exp 
\left[
\rmi \sum_i \hat{\lambda}_i (\alpha - \lambda_i)
s_i \right] \nonumber \\
\mbox{} & \times & 
\exp
{\Biggr [}
-\frac{1}{2K} \sum_{\mu k}(\hat{\lambda}_k^2 (v^\mu -
y^\mu)^2 + \hat{v}_\mu^2 + (\hat{v}^0_\mu)^2 + 2\hat{v}_\mu
\hat{v}^0_\mu s_k s_k^0 \nonumber \\
\mbox{} & + & 2\hat{v}_\mu \hat{\lambda}_k (v^\mu -
y^\mu) s_k + 2\hat{v}_\mu^0 \hat{\lambda}_k (v^\mu - y^\mu) s_k^0
)
{\Biggr ]}.
\end{eqnarray}
By introducing the following order parameters:
\begin{subequations}
\begin{align}
q &=
\frac{1}{K} \sum_{k} s_{k}s_{k}^{0} \\
t &=
\frac{1}{K} \sum_{k}\hat{\lambda}_{k} s_{k} \\
u &=
\frac{1}{K} \sum_{k} \hat{\lambda}_{k}^{2} \\
w &=
\frac{1}{K} \sum_{k} \hat{\lambda}_{k} s_{k}^{0}
\end{align}
\end{subequations}
via
\begin{subequations}
\begin{align}
1 &= \int \frac{\rmd \hat{q}\rmd q}{2\pi/K} 
\exp \left[
\rmi K \hat{q}
\left(q - \frac1K
\sum_k s_k s_k^0
\right)
\right] \\
1 &= \int \frac{\rmd t \rmd \hat{t}}{2\pi/K} 
\exp \left[\rmi K\hat{t}
\left(t -
\frac1K \sum_k \hat{\lambda}_k s_k
\right)
\right] \\
1 &= \int \frac{\rmd u \rmd \hat{u}}{2\pi/K} 
\exp\left[\rmi K \hat{u}
\left(u -
\frac1K \sum_k \hat{\lambda}_k^2
\right)
\right] \\
1 &= \int \frac{\rmd w \rmd \hat{w}}{2\pi/K} 
\exp \left[\rmi K \hat{w}
\left(w-
\frac1K \sum_k \hat{\lambda_k} s_k^0 
\right)
\right]
\end{align}
\end{subequations}
we obtain
\begin{multline}
\bra g_0 \ket = \left(\frac{\beta_s}{2\pi}
\right)^\frac{N}{2}\int \frac{\rmd t \rmd \hat{t}}{2\pi/K}
\frac{\rmd u \rmd \hat{u}}{2\pi/K} \frac{\rmd w \rmd
\hat{w}}{2\pi/K} \frac{\rmd q \rmd \hat{q}}{2\pi/K} \rme^{\rmi K
(\hat{t}t + \hat{u} u + \hat{w} w + \hat{q}q)} \\
\times \exp\Bigg[\alpha K \log \int \rmd y \frac{\rmd v \rmd
\hat{v}}{2\pi} \frac{\rmd v^0 \rmd \hat{v}^0}{2\pi}
\rme^{-\frac{\beta_s}{2} \left(y - v^0 \right)^2 + \rmi (\hat{v}^0
v^0 + \hat{v} v)- \frac12 \hat{v}^2 -\frac12 (\hat{v}^0)^2 -q\hat{v}
\hat{v}^0 -\frac{u}{2}
(v - y)^2} \\
\times \rme^{- t\hat{v}(v - y) -w \hat{v}^0 (v -
y)} \Bigg] \\
\times \exp\Bigg[K \log \frac12 \sum_{s,s^0}\int_0^\infty \rmd \lambda
\int_{-\infty}^{\infty} \frac{\rmd \hat{\lambda}}{2\pi} \rme^{\rmi
\hat{\lambda} (\alpha - \lambda)s - \rmi (\hat{q} s s^0 + \hat{u}
\hat{\lambda}^2 + \hat{t} \hat{\lambda}s + \hat{w} \hat{\lambda}
s^0)}\Bigg]
\label{eq:anneal_g0_3}.
\end{multline}
Now, we rotate the variables $\hat{q},\hat{u}$ to $-\rmi
\hat{q},-\rmi \hat{u}$, and  focus on the part:
\begin{multline}
I=\frac12\sum_{s,s^0}\int_0^\infty \rmd \lambda
\int_{-\infty}^{\infty} \frac{\rmd \hat{\lambda}}{2\pi} \rme^{\rmi
\hat{\lambda} (\alpha - \lambda)s - (\hat{q} s s^0 + \hat{u}
\hat{\lambda}^2 + \rmi\hat{t}s
\hat{\lambda} + \rmi\hat{w} \hat{\lambda}  s^0)} \nonumber \\
= \frac12\sum_{s,s^0} \rme^{-\hat{q} s s^0}
\int_0^\infty \rmd \lambda \int_{-\infty}^{\infty} \frac{\rmd
\hat{\lambda}}{2\pi} \rme^{-\hat{u} \hat{\lambda}^2 + \rmi
\hat{\lambda} [(\alpha - \lambda - \hat{t})s - \hat{w} s^0])}
\end{multline}
By  changing the variable $\hat{\lambda} \to
\hat{\lambda}/\sqrt{2\hat{u}}$, we obtain
\begin{eqnarray}
I = \cosh(\hat{q})
+ \frac12 \left[\rme^{-\hat{q}} \mbox{Erf} \left(\frac{\alpha
-\hat{t} - \hat{w}}{2\sqrt{\hat{u}}} \right) + \rme^{\hat{q}} \mbox{Erf}
\left(\frac{\alpha -\hat{t} + \hat{w}}{2\sqrt{\hat{u}}} \right)
\right].
\end{eqnarray}
where we used $\int Dz\, {\rm e}^{az}={\rm e}^{a^{2}/2}$ and ${\rm
Erf}(z)=(2/\sqrt{\pi}) \int_{0}^{z} dt\, {\rm e}^{-t^{2}}$. We also
calculate the other integral in equation (\ref{eq:anneal_g0_3}),
that is,
\begin{multline}
\int \rmd y \frac{\rmd v \rmd \hat{v}}{2\pi} \frac{\rmd v^0 \rmd
\hat{v}^0}{2\pi} \rme^{-\frac{\beta_s}{2} \left(y - v^0 \right)^2 +
\rmi (\hat{v}^0 v^0 + \hat{v} v)- \frac12 \hat{v}^2 -\frac12
(\hat{v}^0)^2 -q\hat{v} \hat{v}^0 -\frac{u}{2}
(v - y)^2- t\hat{v}(v - y) -w \hat{v}^0 (v - y)} \\
= \int \frac{\rmd y  \rmd v }{\sqrt{2\pi (1 + \beta_s(1-q^2)))}}
\rme^{-\frac{\beta_s y^2}{2[1 + \beta_s(1-q^2)]} -\frac{v^2
q^2\beta_s}{2[1 + \beta_s(1-q^2)]}
-\frac{(qt-w)^2(v-y)^2\beta_s}{2[1 + \beta_s(1-q^2)]}} \nonumber\\
\times \rme^{- \frac{u}{2}
(v - y)^2- \frac12 v^2 - \frac12 t^2(v - y)^2 - tv(v-y)
-\frac{ vq(qt-w)(v-y)\beta_s}{[1 + \beta_s(1-q^2)]} +
\frac{\beta_s y v q}{[1 + \beta_s(1-q^2)]}
+ \frac{\beta_s y(qt-w)(v-y)}{[1 + \beta_s(1-q^2)]}}
\end{multline}
where we rotated the variables $t,w$ to $-\rmi t,-\rmi w$. In this
expression, the coefficients of $y^{2},v^{2}$ and $yv$ are given by,
respectively,
\begin{subequations}
\begin{align}
A &= \frac{\beta_s[1 + qt-w]^2}{[1 +\beta_s(1-q^2)]} +u + t^2 \\
B &= \frac{ \beta_s[q(1+t)-w]^2}{[1 + \beta_s(1-q^2)]} +u +(1+t)^2\\
C & = \frac{[1 + qt-w][q + qt - w]\beta_s}{[1 + \beta_s(1-q^2)]} + u +
t(t+1)
\end{align}
\end{subequations}
Thus, our integral to be calculated is now
written as
\begin{multline}
\sqrt{2\pi} \int \frac{\rmd y \rmd v}{2\pi \sqrt{1 +
\beta_s(1-q^2)}}
\rme^{-\frac{A}{2} y^2 - \frac{B}{2} v^2 + Cyv}
= \frac{1}{\sqrt{(BA - C^2)[1 + \beta_s(1-q^2)]}} \nonumber \\
= \frac{\sqrt{2\pi}}{\sqrt{u[1 + 2\beta_s(1-q)] + \beta_s(1 + t
-w)^2}}.
\end{multline}
So the average number of locally minimum solutions is given by
\begin{multline}
\bra g_0 \ket=\int \frac{\rmd t \rmd \hat{t}}{2\pi/K} \frac{\rmd u
\rmd \hat{u}}{2\pi/K} \frac{\rmd w \rmd \hat{w}}{2\pi/K} \frac{\rmd
q \rmd \hat{q}}{2\pi/K} \rme^{ K \Phi(
t,u,w,q,\hat{t},\hat{u},\hat{w},\hat{q})} \\
\Phi = \hat{t}t + \hat{u} u + \hat{w} w + \hat{q}q
-\frac{\alpha}{2} \log \left\{ u[1 + 2\beta_s(1-q)] + \beta_s(1 + t
-w)^2\right\}\nonumber\\
+\log \left\{\cosh(\hat{q}) + \frac12 \left[\rme^{-\hat{q}}
\mbox{Erf} \left(\frac{\alpha -\hat{t} - \hat{w}}{2\sqrt{\hat{u}}}
\right) + \rme^{\hat{q}} \mbox{Erf} \left(\frac{\alpha -\hat{t} +
\hat{w}}{2\sqrt{\hat{u}}} \right) \right]\right\} + \frac{\alpha}{2}
\log \beta_s
\end{multline}
which is just the expression given in (\ref{eq:anneal_g0_2}).
\section{Evaluation of the replica symmetric saddle point
surface}
In this appendix,
we explain the derivation of
the replica symmetric saddle point, namely,
(\ref{eq:quench_surface_rs})
from (\ref{eq:quench_surface})
with (\ref{eq:g0_n}).
By using the replica
symmetric ansatz :
\begin{subequations}
\begin{align}
m_{\alpha} &= m, \,\,\,\hat{m}_{\alpha}=\hat{m} \\
w_{\alpha} &= w, \,\,\,\hat{w}_{\alpha}=\hat{w} \\
q_{\alpha \beta} &= q,\,\,\,\hat{q}_{\alpha \beta} =\hat{q} \\
u_{\alpha} &= u,\,\,\,\hat{u}_{\alpha}=\hat{u}
\end{align}
\end{subequations}
we can rewrite the saddle point surface $\Phi$ as
\begin{multline}
\Phi_{RS} = \rmi n (\hat{m} m + \hat{w} w + \hat{u}_d u_d +
\hat{r}_d r_d) + \rmi \frac{n(n-1)}{2} (\hat{q} q + \hat{u} u + 2\hat{r}r)
+ \frac{\alpha}{2} \log \frac{\beta_s}{2\pi}\\
+ \log \frac12 \sum_{s^0,s^1,\ldots,s^n} \prod_{\alpha}
\left[\int
\frac{\rmd \lambda^\alpha \rmd \hat{\lambda}^\alpha}{2\pi}
\right]
\exp\left[\rmi \sum_{\alpha} \hat{\lambda}^\alpha (\alpha -
\lambda^\alpha) - \rmi s^0 \sum_\alpha (\hat{m} s^\alpha+
\hat{w} s^\alpha \hat{\lambda}^\alpha)
\right] \\
\exp\left[-\rmi \hat{q} \sum_{\alpha <
\beta} s^\alpha s^\beta  -\rmi \hat{u}_d \sum_{\alpha}
  \hat{\lambda}_\alpha^2 -\rmi \hat{u} \sum_{\alpha < \beta} s^\alpha
s^\beta
\hat{\lambda}^\alpha \hat{\lambda}^\beta - \rmi \hat{r}_d
\sum_{\alpha} \hat{\lambda}^\alpha - \rmi \hat{r} \sum_{\alpha \neq \beta}
s_\alpha  s_\beta \hat{\lambda}^\beta
\right] \\
+ \alpha \log \int \left[\rmd y
\prod_{\alpha = 0}^n \frac{\rmd v^\alpha \rmd \hat{v}^\alpha}{2\pi}
\right]
\exp\left[
-\frac{\beta_s}{2} \left(y - v^0 \right)^2 + \rmi
\sum_{\alpha = 0}^n \hat{v}^\alpha v^\alpha - \frac{1}{2}\hat{v}_0^2 -
\frac12 \sum_{\alpha = 1}^n \hat{v}_\alpha^2 -\frac12 q \sum_{\alpha
\neq\beta}
\hat{v}^\alpha \hat{v}^\beta
\right]\\
\exp {\Biggr [} 
-r_d \sum_{\alpha} \hat{v}_\alpha (v_\alpha - y) - r
  \sum_{\alpha \neq \beta} \hat{v}_\alpha (v_\beta - y) - \frac12 u_d
  \sum_\alpha (v_\alpha - y)^2 \\
  - \frac12 u \sum_{\alpha \neq \beta}
(v_\alpha -
  y)(v_\beta - y) - \hat{v}_0 m \sum_\alpha \hat{v}_\alpha - \hat{v}_0
  w \sum_\alpha (v_\alpha - y)
{\Biggr ]} \\
=  n (\hat{m} m + \rmi \hat{w} w + \hat{u}_d u_d -\rmi
\hat{r}_d r_d) + \frac{n(n-1)}{2} (\hat{u} u - \hat{q} q -2\rmi\hat{r}r)
+ \frac{\alpha}{2} \log \frac{\beta_s}{2\pi}\\
+ \log I_1 + \alpha \log I_2
\label{eq:rs_surface_2}.
\end{multline}
where we made the rotations:
$\rmi \hat{m} \to \hat{m},
\rmi \hat{u} \to \hat{u},
\rmi \hat{q} \to -\hat{q}$ and
$\rmi \hat{r} \to \hat{u}$.
We now focus on the first
integral, $I_{1}$,
\begin{multline}
I_1 = \frac12 \sum_{s^0,s^1,\ldots,s^n} \prod_{\alpha}
\left[\int
\frac{\rmd \lambda^\alpha \rmd \hat{\lambda}^\alpha}{2\pi}
\right]
\exp\left[\rmi \sum_{\alpha} \hat{\lambda}^\alpha (\alpha -
\lambda^\alpha) - \rmi s^0
\sum_\alpha  s^\alpha  (\hat{m}+  \hat{w} \hat{\lambda}^\alpha)
\right] \\
\times \exp{\Biggr [} 
-\rmi\frac{\hat{q}}{2} 
\left(\sum_{\alpha} s^\alpha \right)^2 + \rmi
\frac{\hat{q}n}{2} - \rmi 
\left(\hat{u}_d  - \frac{ \hat{u}}{2}\right)
\sum_{\alpha} \hat{\lambda}_\alpha^2
- \rmi \frac{\hat{u}}{2}
\left(\sum_\alpha s_\alpha \hat{\lambda}_\alpha
\right)^2 \\
- \rmi (\hat{r}_d - \hat{r}) \sum_{\alpha} \hat{\lambda}^\alpha -
\rmi
\hat{r} \sum_{\alpha \beta} s^\alpha s^\beta\hat{\lambda}^\beta
{\Biggr ]}.
\end{multline}
We consider the identity:
\begin{eqnarray}
\rmi \sum_{\alpha \beta} s^\alpha s^\beta \hat{\lambda}^\beta
=\frac12 
\left\{\sum_\alpha(s^\alpha + \rmi s^\beta
\hat{\lambda}^\beta)
\right\}^2 - \frac12
\left(\sum_\alpha s^\alpha
\right)^2 + \frac12
\left(\sum_\alpha s^\alpha \hat{\lambda}^\alpha
\right)^2
\end{eqnarray}
Then, we have
\begin{multline}
I_{1} = \frac12 \sum_{s^0,s^1,\ldots,s^n} 
\exp
\left[-s^0 \sum_\alpha
s^\alpha  \hat{m}
\right] \prod_{\alpha}
\left[\int \frac{\rmd \lambda^\alpha
\rmd \hat{\lambda}^\alpha}{2\pi}
\right] \exp\left[
\rmi \sum_{\alpha}
\hat{\lambda}^\alpha (\alpha + \hat{r}_d - \hat{r} - s^0 s^\alpha
\hat{w} - \lambda^\alpha)
\right] \\
\times 
\exp{\Biggr [}
\frac{\hat{q} - \hat{r}}{2} 
\left(
\sum_{\alpha} s^\alpha
\right)^2 -
\frac{\hat{q}n}{2} - 
\left(
\hat{u}_d  - \frac{\hat{u}}{2}
\right) \sum_{\alpha}
\hat{\lambda}_\alpha^2
- \frac{\hat{u}
- \hat{r}}{2}
\left(
\sum_\alpha s_\alpha \hat{\lambda}_\alpha
\right)^2 \nonumber \\
+  \frac{ \hat{r}}{2}
\left\{\sum_{\alpha}( s^\alpha
+ \rmi s^\alpha \hat{\lambda}^\alpha)
\right\}^2
{\Biggr ]} \nonumber\\
=
\rme^{-\frac{\hat{q}n}{2}} \frac12 \sum_{s^0} \int Dz_1 Dz_2
Dz_3 \Bigg\{\frac12 \sum_s \rme^{-s^0 s \hat{m} + z_1 \sqrt{\hat{q}
- \hat{r}} s + z_3 \sqrt{\hat{r}} s} \Bigg[1 + \\
\mbox{Erf}\Bigg(\frac{ \alpha + \hat{r}_d - \hat{r} - s s^0 w + z_2
\sqrt{\hat{u} - \hat{r}} s + z_3 \sqrt{r} s}{\sqrt{2(u_d - u)}}
\Bigg) \Bigg]\Bigg\}^n
\end{multline}
where we made the transformation: $\hat{u}_{d} \to \hat{u}_{d}/2$
and performed the Hubbard-Stratonovich transformations for ${\exp}
[(\hat{q}-\hat{r})(\sum_{\alpha}s^{\alpha})^{2}/2], {\exp}
[-(\hat{u}-\hat{r})
(\sum_{\alpha}s_{\alpha}\hat{\lambda}_{\alpha})^{2}/2]$ and ${\exp}
[\hat{r}\{\sum_{\alpha} (s^{\alpha}+\rmi
s^{\alpha}\hat{\lambda}^{\alpha})\}^{2}/2]$. We now turn to the
second integral:
\begin{multline}
I_2 = \int \left[\rmd y
\prod_{\alpha = 0}^n \frac{\rmd v^\alpha \rmd \hat{v}^\alpha}{2\pi}
\right] \\
\times \exp\left[-\frac{\beta_s}{2} \left(y - v^0 \right)^2 + \rmi
\sum_{\alpha = 0}^n \hat{v}^\alpha v^\alpha - \frac{1}{2}\hat{v}_0^2 -
\frac12 \sum_{\alpha = 1}^n \hat{v}_\alpha^2 -\frac12 q \sum_{\alpha
\neq\beta}
\hat{v}^\alpha \hat{v}^\beta\right] \\
\times \exp{\Biggr [}
-r_d \sum_{\alpha} \hat{v}_\alpha (v_\alpha - y) - r
  \sum_{\alpha \neq \beta} \hat{v}_\alpha (v_\beta - y) - \frac12 u_d
  \sum_\alpha (v_\alpha - y)^2 \\
  - \frac{u}{2} \sum_{\alpha \neq \beta}
(v_\alpha -
  y)(v_\beta - y) - \hat{v}_0 m \sum_\alpha \hat{v}_\alpha - \hat{v}_0
  w \sum_\alpha (v_\alpha - y)
  {\Biggr ]} \\
= \int \frac{\rmd y \rmd v^0 \rmd \hat{v}^0}{2\pi}
\exp\left[-\frac{\beta_s}{2}(y - v^0)^2 + \rmi v^0 \hat{v}^0 - \frac12
\hat{v}_0^2
\right] \int \left[\prod_{\alpha=1}^n \frac{\rmd v_\alpha
\rmd \hat{v}_\alpha}{2\pi}\right] \\
\times 
\exp\left[
\rmi \sum_\alpha v_\alpha
\hat{v}_\alpha - \hat{v}^0 m\sum_\alpha\hat{v}_\alpha -
\hat{v}^0\sum_\alpha w(v_\alpha- y)\right] \nonumber \\
\times \exp{\Biggr [}- \frac{q-r}{2}
\left(
\sum_\alpha \hat{v}^\alpha
\right)^2 + \frac{q-1}{2} \sum_\alpha
\hat{v}_\alpha^2 
-\frac{u-r}{2}
\left\{\sum_\alpha (v_\alpha - y)
\right\}^2 \\ 
+\frac12(u - u_d) \sum_\alpha (v_\alpha - y)^2 + (r - r_d)
\sum_\alpha \hat{v}_\alpha(v_\alpha - y) - \frac{r}{2}
\left\{\sum_\alpha(
\hat{v}^\alpha + v_\alpha - y)
\right\}^2
{\Biggr ]}.
\end{multline}
The Hubbard-Stratonovich transformations 
with respect to the factors: 
${\exp}[-(q-r)(\sum_{\alpha}\hat{v}_{\alpha})^{2}/2],
{\exp}[-(u-r)\{\sum_{\alpha}(v_{\alpha}-u)\}^{2}/2]$ and
${\exp}[-r\{\sum_{\alpha}
(\hat{v}_{\alpha}+v_{\alpha}-u)\}^{2}/2]$
give
\begin{multline}
I_{2} = \int Dz_1 Dz_2 Dz_3 \frac{\rmd y \rmd v^0 \rmd
\hat{v}^0}{2\pi} \rme^{-\frac{\beta_s}{2}(y - v^0)^2 + \rmi v^0
\hat{v}^0 - \frac12 \hat{v}_0^2} \Bigg\{ \int
\frac{\rmd v\rmd \hat{v}}{2\pi}
\rme^{\rmi v\hat{v} - \hat{v}^0 m\hat{v} - \hat{v}^0 w(v-y)
}\nonumber\\
\rme^{z_1 \sqrt{-(q-r)} \hat{v} + \frac{q-1}{2}\hat{v}^2 +
z_2\sqrt{ -(u-r)}(v - y) + \frac12(u - u_d)(v - y)^2 + (r - r_d)
\hat{v}(v - y)  + z_3 \sqrt{-r}( \hat{v} + v-y)} \Bigg\}^n.
\end{multline}
We should notice that $I_{2}$ now has the form:
\begin{eqnarray}
I_{2} = \bra \bra 1 + n \log X \ket_I \ket_z
\end{eqnarray}
with the averages over the disorder variables ($z_1,z_2,z_3$)
given by $\langle \cdots \rangle_{z}$ and
over the measure on $y,v^0 \hat{v}^0$ given by $\langle \cdots \rangle_{I}$.
We focus on
the inner integral $X$.
By making the change of variable
$v \to v+y$, we have
\begin{multline}
X = \frac{\rme^{-\frac{(y+ \rmi \hat{v}^0 m + z_1 \sqrt{q-r} + z_3
\sqrt{r})^2}{2(1-q)}}}{\sqrt{(1 + \rmi r_d - \rmi r)^2 + (u_d - u)(1 -
q)}}\\
\times \rme^{ \frac{1-q}{2\{(1 + \rmi r_d - \rmi r)^2 + (u_d - u)(1 -
q)\}}(\rmi z_3 \sqrt{r} + \rmi z_2 \sqrt{u - r} - \hat{v}^0 w -
\frac{1 + \rmi r_d - \rmi r}{(1-q)}[y+ \rmi \hat{v}^0 m + z_1\sqrt{q
- r} + z_3 \sqrt{r}])^2}.
\end{multline}
Then, we have
\begin{multline}
\bra\bra \log X \ket_I \ket_z =
\sqrt{\frac{2\pi}{\beta_s}}\Bigg\{ - \frac12 \log[(1 + \rmi r_d -
\rmi r)^2 + (u_d - u)(1 - q)] - \frac{1+ \beta_s^{-1} + q -
2m}{2(1-q)}\nonumber\\
+ \frac{1-q}{2\{(1 + \rmi r_d - \rmi r)^2 + (u_d - u)(1 - q)\}}
\Bigg[ -u + 2\left(\frac{1 + \rmi r_d - \rmi r)}{1-q}\right)(\rmi w
- \rmi
r ) \nonumber\\
+ \left( \frac{1 + \rmi r_d - \rmi r}{1-q} \right)^2 (1 +
\beta_s^{-1} + q -2m ) \Bigg]\Bigg\}\nonumber
\end{multline}
where we used the fact:
\begin{eqnarray}
\bra 1 \ket_I = \bra v_0^2 \ket_I = \bra y v_0 \ket_I =
\sqrt{\frac{2\pi}{\beta_s}} \qquad \bra y^2 \ket_I =
\sqrt{\frac{2\pi}{\beta_s}}(1 + \beta_s^{-1})\\
\bra v_0 \hat{v}_0 \ket_I = \bra y \hat{v}_0 \ket_I = \rmi
\sqrt{\frac{2\pi}{\beta_s}} \qquad \bra y \ket_I = \bra \hat{v}_0
\ket_I = \bra v_0 \ket_I = \bra \hat{v}_0^2 \ket_I = 0
\end{eqnarray}
with the definition:
\begin{eqnarray}
\bra \ldots \ket_{I} = \int \frac{\rmd y \rmd v_0 \rmd
\hat{v}_0}{2\pi} \rme^{-\frac{\beta_s}{2}(y - v_0)^2 - \frac12
\hat{v}_0^2 + \rmi v_0 \hat{v}_0} (\ldots).
\end{eqnarray}
By rotating the variables
$\rmi w \to w,
\rmi r \to r$, we obtain
\begin{multline}
\bra\bra \log X \ket_I \ket_z =
\sqrt{\frac{2\pi}{\beta_s}}\Bigg\{ - \frac12 \log[(1 + r_d - r)^2 +
(u_d - u)(1 - q)] \nonumber\\
+ \frac{2(w-r)(1 + r_d - r)
-u(1-q) -(1 + \beta_s^{-1} + q - 2m)(u_d - u) }{2\{(1 + r_d - r)^2 +
(u_d - u)(1 - q)\}} \Bigg\}.
\end{multline}
Putting  $I_{1}$ and $I_{2}$
into (\ref{eq:rs_surface_2}), we obtain the replica symmetric
saddle point surface:
\begin{multline}
\frac{1}{n} \Phi_{RS} = \hat{m}m+ \hat{w} w + \frac{1}{2}
\hat{q} q + \frac12 \hat{u}_d u_d - \frac{1}{2}
\hat{u}u - \hat{r}_d r_d + \hat{r}r\\
-\frac{\hat{q}}{2}  + \frac12 \sum_{s^0} \int Dz_1 Dz_2 Dz_3
\log \Bigg\{\frac12 \sum_s \rme^{-s^0 s \hat{m} + z_1 \sqrt{\hat{q}
- \hat{r}} s + z_3 \sqrt{\hat{r}} s} \Bigg[1  \\
+ \mbox{Erf}\Bigg(\frac{ \alpha + \hat{r}_d  - \hat{r} - s s^0
\hat{w} + z_2 \sqrt{\hat{u} - \hat{r}} s + z_3 \sqrt{\hat{r}}
s}{\sqrt{2(\hat{u}_d - \hat{u})}}
\Bigg) \Bigg]\Bigg\}\\
+ \alpha \Bigg\{ - \frac12 \log[(1 + r_d - r)^2 +
(u_d - u)(1 - q)] \nonumber\\
+ \frac{2(w-r)(1 + r_d - r)
-u(1-q) -(1 + \beta_s^{-1} + q - 2m)(u_d - u) }{2\{(1 + r_d - r)^2 +
(u_d - u)(1 - q)\}} \Bigg\}.
\end{multline}


\begin{thebibliography}{9}

\bibitem{Nishimori} H. Nishimori:
\textit{Statistical Physics of
Spin Glass and Information
Processing} (Oxford University Press, 2001).

\bibitem{Tanaka1}
T. Tanaka,
Europhys. Lett. {\bf 54} (4), 540 (2001).

\bibitem{Tanaka2}
T. Tanaka,
IEEE Trans. on Information Theory
{\bf 48} (11), 2888 (2002).


\bibitem{TanakaOkada}
T. Tanaka and M. Okada, IEEE Trans. on Information Theory {\bf 51}
(2), 700 (2005).

\bibitem{Kabashima}
Y. Kabashima, J. Phys. A: Math. Gen. {\bf 36} 11111 (2003)

\bibitem{MimuraOkada}
K. Mimura and M. Okada, {J. Phys. A: Math. Gen} {\bf 38} 9917 (2005)

\bibitem{HatchettOkada}
J.P.L. Hatchett and M. Okada {J. Phys. A: Math. Gen} (in press)

\bibitem{TE}
F. Tanaka and S.F. Edwards,
J. Phys. F : Metal Phys.
{\bf 10} 2769 (1980).

\bibitem{TE2}
F. Tanaka and S.F. Edwards,
J. Phys. F : Metal Phys.
{\bf 10} 2779 (1980).


\bibitem{SK}
D. Sherrington and
S. Kirkpatrick,
Phys. Rev. Lett.
{\bf 35}, 1792 (1975).


\end{thebibliography}
\end{document}